\documentclass[11pt,preprint]{aastex}

\begin{document}

\shorttitle{Disk Structures in Ophiuchus}

\shortauthors{Andrews et al.}

\title{Protoplanetary Disk Structures in Ophiuchus II: Extension to Fainter Sources}

\author{Sean M. Andrews\altaffilmark{1,2}, D. J. Wilner\altaffilmark{1}, A. M. Hughes\altaffilmark{1}, Chunhua Qi\altaffilmark{1}, and C. P. Dullemond\altaffilmark{3}}

\email{sandrews@cfa.harvard.edu}

\altaffiltext{1}{Harvard-Smithsonian Center for Astrophysics, 60 Garden Street, Cambridge, MA 02138}
\altaffiltext{2}{Hubble Fellow}
\altaffiltext{3}{Max Planck Institut f\"{u}r Astronomie, K\"{o}nigstuhl 17, 69117 Heidelberg, Germany}

\begin{abstract}
We present new results from a significant extension of our previous high 
angular resolution (0\farcs3 $\approx$ 40\,AU) Submillimeter Array survey of 
the 340\,GHz (880\,$\mu$m) thermal continuum emission from dusty circumstellar 
disks in the $\sim$1\,Myr-old Ophiuchus star-forming region.  An expanded 
sample is constructed to probe disk structures that emit significantly lower 
millimeter luminosities (hence dust masses), down to the median value for T 
Tauri stars.  Using a Monte Carlo radiative transfer code, the millimeter 
visibilities and broadband spectral energy distribution for each disk are 
simultaneously reproduced with a two-dimensional parametric model for a viscous 
accretion disk that has a surface density profile $\Sigma \propto 
(R/R_c)^{-\gamma} \exp{[-(R/R_c)^{2-\gamma}]}$.  We find wide ranges of 
characteristic radii ($R_c = 14$-198\,AU) and disk masses ($M_d = 
0.004$-0.143\,M$_{\odot}$), but a narrow distribution of surface density 
gradients ($\gamma = 0.4$-1.1) that is consistent with a uniform value $\langle 
\gamma \rangle = 0.9\pm0.2$ and independent of mass (or millimeter 
luminosity).  In this sample, we find a correlation between the disk 
luminosity/mass and characteristic radius, such that fainter disks are both 
smaller and less massive.  We suggest that this relationship is an imprint of 
the initial conditions inherited by the disks at their formation epoch, compare 
their angular momenta with those of molecular cloud cores, and speculate on how 
future observations can help constrain the distribution of viscous evolution 
timescales.  No other correlations between disk and star properties are found.  
The inferred disk structures are briefly compared with theoretical models for 
giant planet formation, although resolution limitations do not permit us to 
directly comment on material inside $R \approx 20$\,AU.  However, there is some 
compelling evidence for the evolution of dust in the planet formation region: 
4/17 disks in the sample show resolved regions of significantly reduced 
millimeter optical depths within $\sim$20-40\,AU of their central stars.  
\end{abstract}

\section{Introduction}

Direct observations of the reservoirs of planet-building material $-$ the disks 
around young stars $-$ should play a critical role in developing theoretical 
models of planet formation.  Ultimately, the goal of those models is to 
elucidate the physical processes involved in making planets by incorporating 
the observed properties of circumstellar disks and successfully reproducing the 
demographic characteristics of the planets in our Solar System and those around 
other stars \citep[e.g.,][]{ida04,mordasini09}.  Regardless of the favored 
mechanism, two basic requirements for forming planets in a circumstellar disk 
must be satisfied: there must be enough material (gas and dust) in the right 
locations and a sufficient amount of time for the formation mechanism to 
operate \citep[e.g.,][]{pollack96,boss97}.  The former criterion amounts to a 
density threshold, suggesting that observational constraints on the spatial 
distribution of mass in young circumstellar disks are fundamental in 
constructing an empirical foundation for planet formation models.  Moreover, 
those same estimates of disk densities can be used to help characterize the 
viscous accretion process that determines how disk structures evolve 
\citep[e.g.,][]{hartmann98}.

There are some significant observational obstacles to direct measurements of 
disk densities.  Most of the material in these disks is ``dark," composed of 
cold molecular hydrogen and not readily detectable.  Their mass contents must 
be inferred from trace species, particularly from the dust grains that dominate 
the disk opacity.  At radio wavelengths, the thermal continuum emission from 
these dust grains is optically thin and therefore provides an unique probe of 
mass in the disk midplane \citep{beckwith90}.  If that emission can be 
resolved, properly interpreted with radiative transfer calculations, and 
assigned some nominal gas-to-dust mass ratio, it can be used to reconstruct the 
spatial distribution of disk densities.  Interferometric observations of dust 
and trace gas species have helped quantify densities in the outer parts of 
circumstellar disks \citep{kitamura02,aw07,hughes08,isella09}, clearly 
established the presence of vertical temperature and density gradients 
\citep{dartois03,pietu07}, and even identified disks with large central 
cavities that exhibit very little dust emission 
\citep{pietu06,hughes07,hughes09,brown08,brown09,isella10}.  

In a previous study \citep[][hereafter Paper I]{andrews09}, we presented the 
initial part of a high angular resolution (0\farcs3 $\approx 40$\,AU) survey of the 880\,$\mu$m thermal dust emission from protoplanetary disks in the nearby 
Ophiuchus star-forming region.  To take advantage of these observations that 
are sensitive to the mass content inside the planet formation zone ($R \le 
40$\,AU), we developed a radiative transfer modeling toolkit to extract the 
two-dimensional temperature and density structures of these disks through a 
combined fit of their millimeter continuum visibilities and broadband spectral 
energy distributions (SEDs).  Through that modeling effort, we concluded that 
the disks had densities comparable to those expected for the outer parts of the 
primordial disk around the Sun and very similar radial density gradients.  The 
inferred density gradients indicate that whatever mechanism is responsible for 
generating viscosity in these disks acts with a nearly linear radial 
distribution (i.e., the viscosities vary as $\nu \propto R^{\gamma}$, with a 
median $\gamma \approx 0.9$).  Moreover, we resolved large regions ($R \approx 
20$-40\,AU) with very little millimeter emission in the centers of several 
disks, and speculated that these may be the signposts of young planetary 
systems.  However, that initial sample was biased towards the targets that are 
exceptionally bright at millimeter wavelengths, and therefore considerably more 
massive than the typical disk in the Ophiuchus region \citep[see][]{aw07b}.  In 
this article, we extend the survey to double the sample and probe disk 
structures with millimeter luminosities down to the median value for 
$\sim$1\,Myr-old T Tauri stars.  In \S 2 we describe the sample selection 
criteria, new observations, and data calibration.  A brief review of the 
radiative transfer calculations is provided in \S 3, with the modeling results 
for this extension of the sample highlighted in \S 4.  In \S 5, we synthesize 
the disk structure constraints for the full sample in the contexts of the 
viscous evolution process and the prospects for planet formation.  The key 
conclusions from the survey are summarized in \S 6.

\section{Sample Selection, Observations, and Data Reduction}

\begin{figure}
\epsscale{0.6}
\plotone{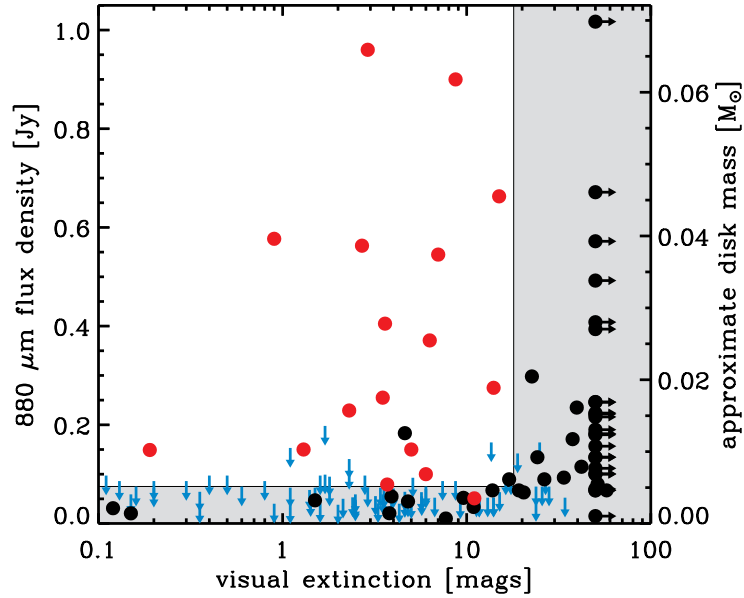}
\figcaption{The millimeter continuum flux densities and visual extinctions for 
young stars in the Ophiuchus molecular clouds (Andrews \& Williams 2007a,b; 
Cieza et al.~2010; additional $A_V$ estimates from Furlan et al.~2009).  The 
targets selected for this sample are shown in red (880\,$\mu$m flux densities 
are from Table \ref{data_table} and Paper I), and the regions that do not meet 
the selection criteria are shaded grey.  Blue arrows represent 3\,$\sigma$ 
upper limits and black points mark sources that have not been observed with the 
SMA at high angular resolution.  Embedded sources with uncertain spectral 
classifications are represented with a lower limit of $A_V = 50$.  When 
880\,$\mu$m flux densities are not available, 1.3\,mm data were scaled up by a 
factor of $\sim$2.2 (see text); the right-hand ordinate axis represents an 
approximate conversion of the flux density scale to total disk masses (see 
Andrews \& Williams 2005; 2007b).  \label{sample}}
\end{figure}

The sample of disk targets for this survey was selected primarily for 
observational and analytical convenience.  To ensure that the SMA data would be 
sufficiently sensitive to probe emission over a large range of spatial scales, 
targets were required to have integrated 880\,$\mu$m flux densities larger than 
$\sim$75\,mJy.  Candidates were selected based on the single-dish photometry 
compiled by \citet{aw07b}.  When no 880\,$\mu$m data were available, we relied 
on the 1.3\,mm flux densities provided by previous surveys 
\citep{andre94,nurnberger98} and conservatively scaled up by the square of the 
wavelength ratio \citep[$\sim$2.2; see][]{aw05,aw07b}.  After this initial cut, 
we excluded targets that lacked sufficient information about their central 
stars.  Without stellar temperatures and luminosities, it is not possbile to 
interpret the observed millimeter data in detail (see \S 3).  In practice, this 
second criterion amounted to an extinction threshold $A_V \lesssim 15$, as 
spectral classifications and luminosity estimates are rare and uncertain for 
more deeply embedded sources \citep{luhman99,wilking05,furlan09}.  The sample 
selection-space set by these criteria is shown in Figure \ref{sample}, with red 
points marking the selected target disks.  The 880\,$\mu$m flux densities for 
those points correspond to the values derived from the SMA observations 
described here or in Paper I.  One selected target, WL 18, falls below the 
designated millimeter emission threshold: previous 1.3\,mm data led us to 
expect a larger 880\,$\mu$m flux density than was observed.  One other target, 
RX J1633.9$-$2422, meets the selection criteria but was not observed: its 
millimeter flux density was only recently published \citep{cieza10}. 

These combined criteria yielded 17 disk targets for the sample.  The 
880\,$\mu$m flux density criterion corresponds to the median value found in 
both the Ophiuchus and Taurus star-forming regions \citep{aw05,aw07b}.  
Therefore, the sample defined here fully spans the upper half of the millimeter 
continuum luminosity distribution $-$ or equivalently the circumstellar dust 
mass distribution $-$ for the disks around $\sim$1\,Myr-old stars.  For the 
typical assumptions in converting this emission to total disk masses, this 
limit corresponds to a few Jupiter masses ($\sim$0.004\,M$_{\odot}$) of gas and 
dust, suggesting that the sample is representative of the disk population that 
might eventually be able to produce giant planets.  Moreover, the sample 
includes targets that essentially span the full range of T Tauri star 
properties, including spectral types from M4 to G3, stellar masses of 
0.3-2.0\,M$_{\odot}$, and accretion rates of 
$\sim$10$^{-9}$-10$^{-7}$\,M$_{\odot}$ yr$^{-1}$.

To complete this sample and expand on the initial 9 disks discussed in Paper I, 
8 additional targets were observed with the very extended (V; 8-509\,m 
baselines) and compact (C: 6-70\,m baselines) configurations of the 
Submillimeter Array interferometer \citep[SMA;][]{ho04} in 2009 March and May.  
A journal of these SMA observations is provided in Table \ref{obs_journal}.  
The SMA double sideband receivers were tuned to a local oscillator (LO) 
frequency of 340.755\,GHz (880\,$\mu$m).  Each sideband was divided into 24 
partially overlapping 104\,MHz chunks centered $\pm$5\,GHz from the LO 
frequency.  The central chunk in the upper sideband was sampled at a factor of 
4 higher spectral resolution than the others, in an effort to observe the CO 
$J$=3$-$2 transition (345.796\,GHz) in 0.70\,km s$^{-1}$ channels.  The 
observing sequence interleaved disk targets with nearby quasars, J1625$-$254 
and J1626$-$298, in an alternating pattern with a total cycle time of 
$\sim$10-15 minutes.  When the targets were at low elevations ($<$20\degr), 
planets (Uranus, Saturn), satellites (Titan, Callisto), and bright quasars (3C 
454.3, 3C 279) were observed as bandpass and absolute flux calibrators 
depending on their availability and the array configuration.  The observing 
conditions were generally very good, with atmospheric opacities $<$0.1 at 
225\,GHz (corresponding to $<$2.0\,mm of precipitable water vapor).  

The data were edited and calibrated as in Paper I using the {\tt MIR} software 
package.  The bandpass response was calibrated with observations of a bright 
planet or quasar, and broadband continuum channels in each sideband were 
generated by averaging the central 82\,MHz in all of the chunks except the one 
reserved for the CO $J$=3$-$2 line.  The visibility amplitude scale was set 
based on observations of planets or satellites and routinely-monitored quasars: 
the typical systematic uncertainty in the absolute flux scale is $\sim$10\%.  
The antenna-based complex gain response of the system as a function of time was 
determined with reference to J1625$-$254, which lies only $\sim$1\degr\ from 
the target disks.  The other quasar in the observing cycle provides a check on 
the quality of the phase transfer in the gain calibration process.  The 
millimeter ``seeing" generated by atmospheric phase noise and any small 
baseline errors is small, 0.1-0\farcs2.  After combining all of the data for 
each target, the standard tasks of Fourier inverting the visibilities, 
deconvolution with the {\tt CLEAN} algorithm, and restoration with a 
synthesized beam were conducted with the {\tt MIRIAD} software package.  High 
resolution maps of the continuum emission were created with a Briggs robust = 
0.2-0.7 weighting scheme for the visibilities, and maps of the CO $J$=3$-$2 
line emission were made with natural weighting for the compact array data 
only.  The relevant data properties from these synthesized maps are compiled in 
Table \ref{data_table}.

The synthesized continuum maps for these targets are featured in Figure 
\ref{images}.  Each high angular resolution map covers 4\arcsec\ on a side, 
corresponding to 500\,AU at the adopted distance of 125\,pc to the Ophiuchus 
clouds \citep{degeus89,knude98,lombardi08,loinard08}.  A centroid position and 
initial estimate of the viewing geometry of the disk $-$ characterized by the 
inclination ($i$) and major axis position angle (PA) $-$ were determined by 
fitting the visibilities with an elliptical Gaussian brightness distribution.  
Because the continuum emission from the SR 24 system is not centrally peaked, 
its centroid position and viewing geometry were estimated by inspection of the 
lower resolution (compact SMA configuration) data only.  In all cases, the CO 
line emission from the disks is significantly contaminated by the local 
molecular cloud environment, and will not be discussed further.

\section{Modeling the Disk Structures}

To interpret these observations in the context of the disk structures $-$ 
physical conditions and key size scales $-$ present in this sample, we followed 
the modeling formalism introduced and discussed in detail in Paper I.  To 
briefly summarize that procedure, we first adopt a parametric prescription for 
a two-dimensional, flared density structure based on a simple model for the 
viscous evolution of an accretion disk \citep{lyndenbell74,hartmann98}.  The 
model assumes an anomalous disk viscosity that varies with radius as $\nu 
\propto R^{\gamma}$ and has a surface density profile
\begin{equation}
\Sigma = (2-\gamma) \frac{M_d}{2 \pi R_c^2} \left( \frac{R}{R_c} \right)^{-\gamma} \exp{\left[-\left(\frac{R}{R_c}\right)^{2-\gamma}\right]},
\end{equation}
where $R_c$ is a characteristic scaling radius and $M_d$ is the disk mass.  At 
a given radius, that column density is vertically distributed so that it falls 
off with altitude above the midplane like a Gaussian with scale-height $H 
\propto R^{1+\psi}$.  Once a density structure has been specified with a set of 
five parameters, \{$M_d$, $\gamma$, $R_c$, $H_c$, $\psi$\} (where $H_c$ is the 
scale-height at the characteristic radius), it is populated with a spatially 
homogeneous size distribution of dust grains.  As described in Paper I, we 
consider dust with an interstellar medium composition 
\citep{draine84,weingartner01} and a power-law distribution of sizes (with an 
index of -3.5) between 0.005\,$\mu$m and 1\,mm.  This dust structure is then 
irradiated by a central star of fixed temperature and luminosity to determine 
an internally-consistent disk temperature structure, using the two-dimensional 
Monte Carlo radiative transfer code {\tt RADMC} \citep{dullemond04a}.  The 
results of the Monte Carlo simulation are then coupled with a post-processing 
raytracing code to compute a synthetic dataset for a given viewing geometry, 
consisting of both a broadband spectral energy distribution (SED) and a set of 
millimeter continuum visibilities that sample the Fourier plane in the same way 
as the SMA data.  These synthetic data are simultaneously compared with the 
observations, and the process is iterated until the fit converges on a minimum 
joint $\chi^2$ value (see Paper I for details).  

The stellar properties used to generate the Kurucz spectra that irradiate the 
disk structures are listed in Table \ref{stars_table}, based on both literature 
measurements and matches to the optical/near-infrared SED.  Because the 
spatio-kinematic information from the CO line emission in these disks is 
sufficiently contaminated by the local cloud material, estimates of the disk 
inclinations and position angles were determined solely from the elliptical 
Gaussian fits to the continuum emission described in \S 2.  Unfortunately, the 
relatively faint continuum signals from some of these targets leads to 
considerable uncertainty in these viewing geometry estimates.  These are still 
the best values available, but we caution that some systematic uncertainty will 
remain for the density parameters until better emission line data are available 
for these targets.  With such a limited amount of spatially resolved 
information about the two components in the DoAr 24 E system, a detailed 
modeling of this source is beyond the scope of this study.  Some important 
modifications to the modeling process were made for the SR 24 S disk, with the 
details provided in \S 4.2.

\section{Results}

\subsection{Disk Structures}

\begin{figure}
\epsscale{1.0}
\plotone{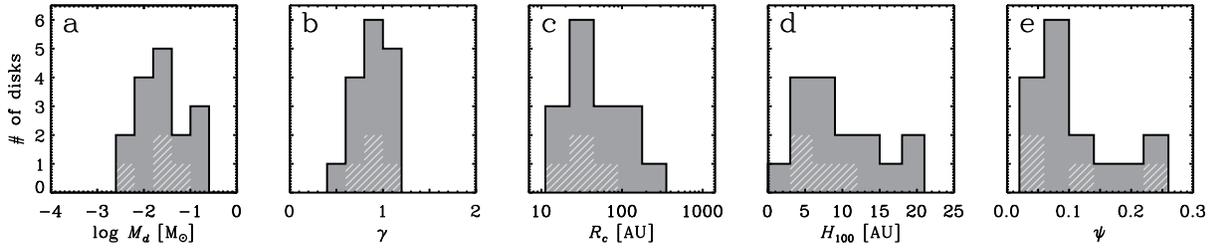}
\figcaption{The derived distributions of the disk structure parameters for the 
composite sample (combining the results presented here and in Paper I).  From 
left to right are the disk masses ($M_d$), radial surface density gradients 
($\gamma$), characteristic radii ($R_c$), scale-heights at 100\,AU ($H_{100}$), 
and the radial scale-height gradients ($\psi$).  The contributions of the four 
disks with diminished millimeter emission in their central regions are hatched. 
\label{histograms}}
\end{figure}

The model parameter values that best reproduce the observations for the full 
sample are compiled in Tables \ref{structure_table} and \ref{trans_table} for 
the disks with continuous emission distributions and those with central 
emission cavities, respectively.  For the continuous disks, we include the 
(fixed) inner disk radii inferred from a simple sublimation argument (see Paper 
I; Table \ref{structure_table}, column 7).  Columns 7 and 8 in Table 
\ref{trans_table} list the estimated sizes ($R_{\rm cav}$) and density 
contrasts ($\delta_{\rm cav}$) used to account for the disks with central 
emission cavities.  Both tables also include the adopted values for the disk 
inclination and position angle, as well as the reduced $\chi^2$ statistics for 
the fits to the SED and visibility datasets separately.  The distributions of 
these values for each parameter are shown together in Figure \ref{histograms}; 
hatched regions mark contributions from the disks with central cavities (around 
SR 24 S, SR 21, DoAr 44, and WSB 60).  In Figure \ref{results_new}, the new 
observations presented here are directly compared with the synthetic datasets 
generated from these best-fit models (see Paper I for the other sample disks).  
From left to right, we display the observed SMA millimeter continuum image (as 
in Figure \ref{images}), the synthesized model image, the imaged residuals, the 
broadband SED, and the elliptically averaged millimeter visibility profile (see 
Paper I for details).  The latter two panels have the best-fit model behavior 
overlaid in red, and the SED panel also shows the input stellar spectrum as a 
dashed blue curve.  Because of its large central emission cavity, the modeling 
results for the SR 24 S disk are shown separately in Figure 
\ref{results_trans}, with inset images synthesized at higher angular resolution 
to facilitate a more detailed comparison.

\begin{figure}
\epsscale{0.6}
\plotone{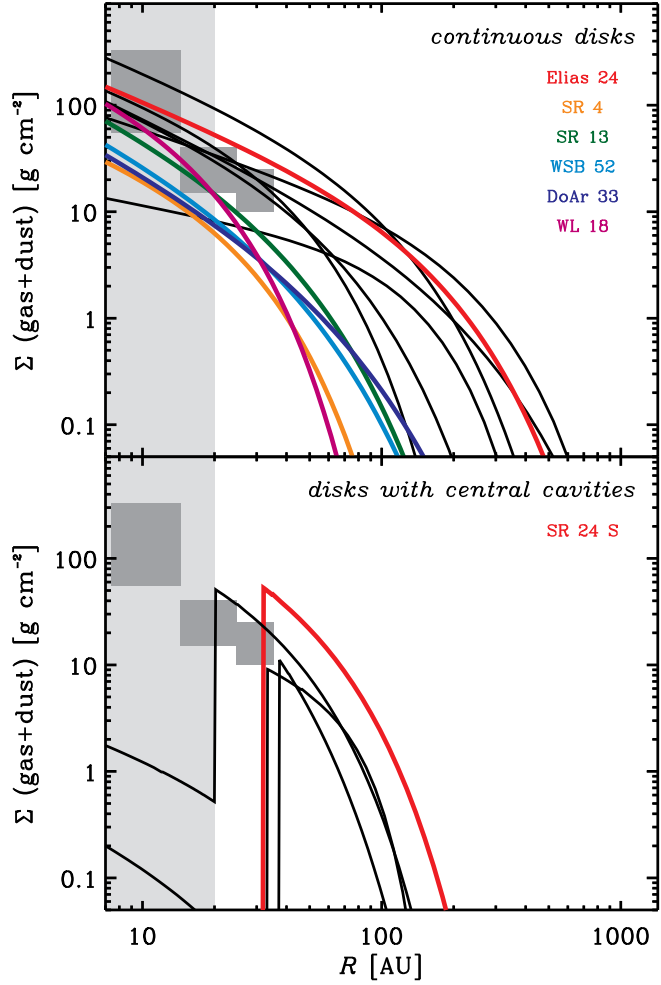}
\figcaption{Surface density profiles for the sample, where the targets with 
continuous emission distributions ({\it top}) and central emission cavities 
({\it bottom}) are separated for clarity.  The new model results presented here 
are highlighted in color, while those from Paper I are shown in black.  The 
light grey band out to a radius of 20\,AU marks the resolution limit of the 
survey.  Dark gray boxes represent the surface densities extrapolated for 
Saturn, Uranus, and Neptune in the standard Minimum Mass Solar Nebula 
\citep{weidenschilling77}.  The density profiles inferred from dust tracers 
have been scaled up assuming a gas-to-dust mass ratio of 100:1.  \label{sigma}}
\end{figure}

The modeling uncertainties and relationships between the observations and model 
parameters were already discussed in detail in Paper I, so we will not repeat 
that effort here.  Instead, we focus on how this expanded sample enables new 
constraints on the spatial distribution of mass in $\sim$1\,Myr-old 
circumstellar disks.  The inferred surface density profiles for the full sample 
are shown together in Figure \ref{sigma}, with the targets presented for the 
first time here highlighted in color.  The disks with large central emission 
cavities are shown in a separate panel.  The light gray band inside 20\,AU 
marks the survey resolution limit, while the dark gray boxes are reference 
points representing the minimal surface densities (with uncertainties) and 
presumed feeding zones for Saturn, Uranus, and Neptune in the canonical model 
for the primordial solar disk \citep{weidenschilling77}.  Although the disks in 
this sample exhibit a wide range of masses ($M_d$; Figure \ref{histograms}a) 
and characteristic radii ($R_c$; Figure \ref{histograms}c), we find a very 
narrow distribution of values for $\gamma$, the radial gradient of the surface 
density profile.  As shown in Figure \ref{histograms}b, that distribution is 
roughly normal with a peak at $\gamma = 0.9$ and a standard deviation of 0.2.  
Since that width is comparable to the modeling uncertainties on $\gamma$ for an 
individual source ($\sim$0.2-0.3; Paper I), the modeling results are consistent 
with a uniform $\gamma$ value being representative of the entire 
sample.\footnote{The only major outlier, the disk around AS 209, has a 
particularly uncertain $\gamma$ estimate due to a poorly-sampled infrared SED 
(see Paper I for details).}  Moreover, by extending the survey to cover much 
fainter targets, we have confirmed that the shape of the surface density 
profile does not significantly change over a wide range of millimeter 
luminosities (or disk masses). 

\begin{figure}
\epsscale{0.6}
\plotone{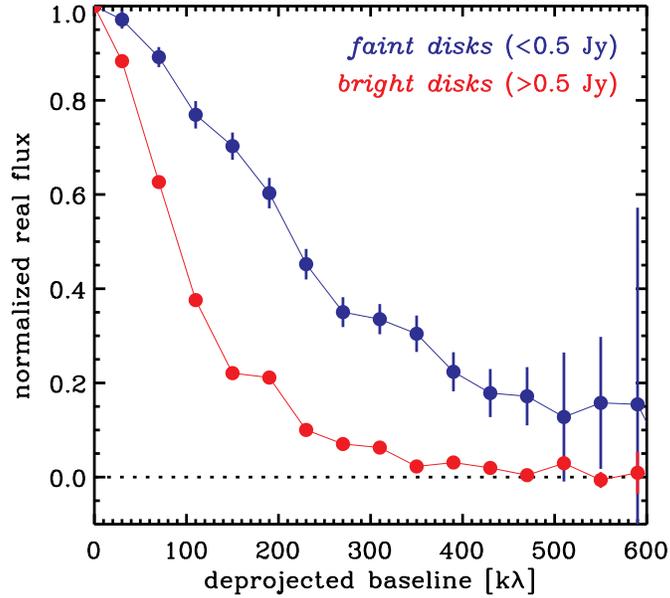}
\figcaption{A direct comparison of the sample-averaged visibility profiles for 
the subsets of disks with integrated 880\,$\mu$m flux densities fainter ({\it 
blue}) and brighter ({\it red}) than 0.5\,Jy.  Each profile was constructed by 
averaging the visibilities for individual disks in annular bins after their 
deprojection according to viewing geometry and normalization by their total 
flux densities.  The disks with large central emission cavities were excluded 
for clarity.  Error bars correspond to the standard deviation in each bin.  The 
significantly larger amount of correlated emission on all spatial scales for 
the fainter subset of disks demonstrates that they are less well resolved, and 
therefore substantially smaller than their brighter counterparts.  This 
empirical relationship is reinforced by the radiative transfer modeling, which 
implies a modest correlation between $M_d$ and $R_c$ (see \S 5.1). 
\label{median_vis}}
\end{figure}

Despite the similar radial density gradients in these disks, the observations 
clearly show a wide variety of millimeter emission morphologies.  Some of that 
diversity is related to how the disks are heated by their central stars, a 
natural outcome of the assortment of stellar properties and vertical 
distributions of dust present in the sample (see Paper I for details).  
However, the density profiles plotted in Figure \ref{sigma} demonstrate that a 
significant range of characteristic radii and masses are also partly 
responsible.  A close examination of those $\Sigma$ profiles reveals that the 
brighter disks at millimeter wavelengths tend to have both higher masses and 
larger characteristic radii.  The former is no surprise, since the vast 
majority of the disk volume at these wavelengths is optically thin, but there 
is no {\it a priori} reason to expect the disk sizes to be correspondingly 
smaller for fainter sources.  This relationship is not a modeling artifact, as 
there is an analogous empirical correlation between the brightness of the 
millimeter emission and how well that emission is resolved.  Figure 
\ref{median_vis} demonstrates this explicitly by comparing the average 
880\,$\mu$m visibility profiles for disks that are brighter ({\it red}) or 
fainter ({\it blue}) than 0.5\,Jy.  These profiles were calculated by averaging 
the visibilities for each subset of disks into annular bins, after deprojecting 
the visibilities for individual disks according to their viewing geometries and 
normalizing their real (correlated) fluxes by dividing off their integrated 
flux densities.  This comparison confirms that the millimeter emission from the 
brighter disks in this sample is more resolved by our SMA data, exhibiting 
substantially less correlated flux on essentially all spatial scales.  Although 
the physical origins of this empirical relationship are not obvious, some 
speculations about its significance for understanding the viscous evolution 
process in these disks are presented in \S 5.

In an effort to identify any trends among the disk structures or the 
characteristics of their stellar hosts, we performed a principle component 
analysis on a subset of such properties for the 12 disks with continuous 
density distributions in this sample.  The analysis included the 5 free disk 
structure parameters, \{$M_d$, $\gamma$, $R_c$, $H_{100}$, $\psi$\}, the 
stellar properties \{$T_\mathrm{eff}$, $L_*$\}, and the accretion rates, 
$\dot{M}_*$ (see Tables \ref{stars_table}, \ref{structure_table}, and 
\ref{viscous_table}).  Due to the additional uncertainties of the stellar 
evolution models used to infer them, we did not include stellar ages, masses, 
or radii directly (although those values are well-correlated with $L_{\ast}$ 
and $T_{\rm eff}$).  As alluded to above, we identified one statistically 
significant correlation (3.3\,$\sigma$; correlation coefficient of 0.85) 
between $M_d$ and $R_c$, which will be discussed in detail in \S 5.1.  The 
first eigenvector from the principle component analysis, accounting for 40\%\ 
of the variance in the data, is dominated by a positive trend relating $R_c$, 
$M_d$, $L_*$, and $\dot{M}_*$.  The trend indicates that the more massive disks 
in this sample are larger and orbiting more luminous - perhaps younger - stars 
that are accreting disk material at higher rates.  While these relationships 
may hint at crucial information related to the viscous evolution process, a 
larger sample will be required to make any definitive conclusions.

\subsection{Commentary on Individual Disks}

\begin{figure}
\epsscale{1.0}
\plotone{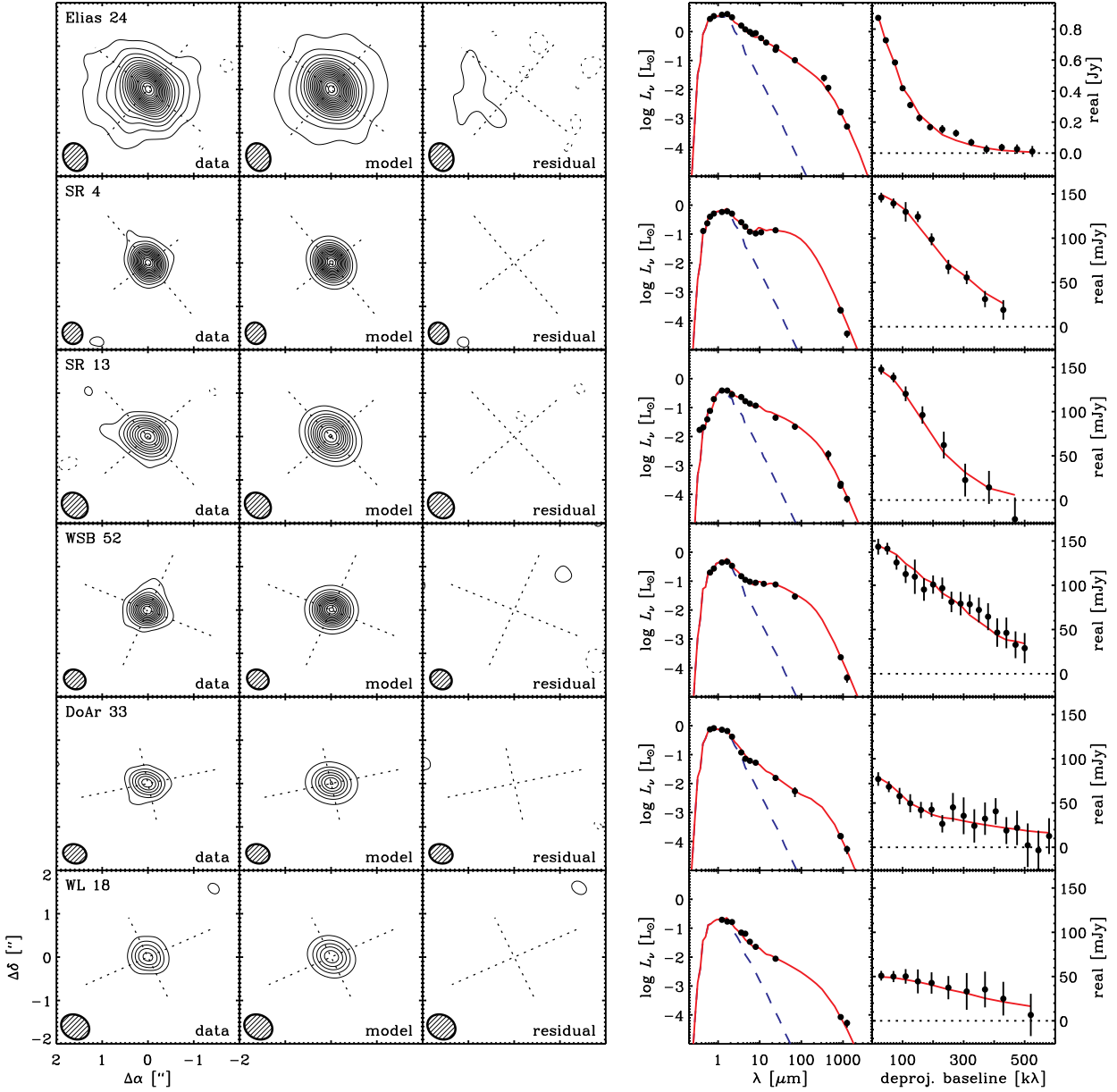}
\figcaption{Comparison of the data with the best-fit disk structure models.  
The left panels show the SMA continuum image, corresponding model image, and 
imaged residuals (data$-$model).  Contours are drawn at the same 3\,$\sigma$ 
intervals in each panel.  Crosshairs mark the disk centers and major axis 
position angle; their relative lengths represent the disk inclination.  The 
right panels show the broadband SEDs and deprojected visibility profiles, with 
best-fit models overlaid in red.  The input stellar photospheres are shown as 
blue dashed curves.  \label{results_new}}
\end{figure}

\begin{figure}
\epsscale{1.0}
\plotone{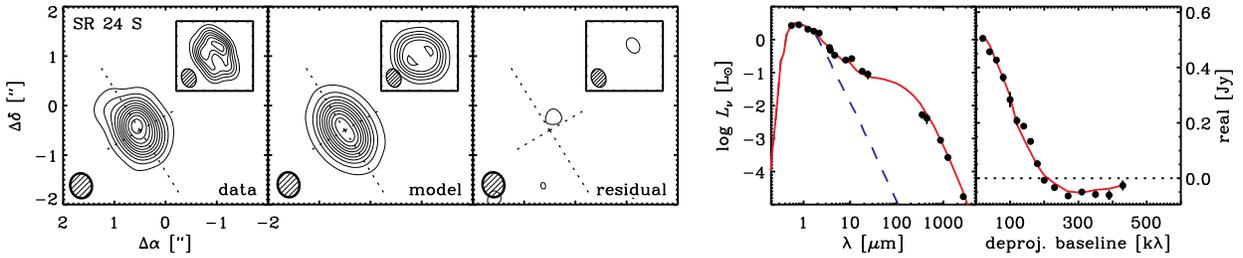}
\figcaption{Same as Figure \ref{results_new} for the SR 24 S disk, which 
features a resolved central emission cavity.  The modifications to the standard 
disk model made for this source in particular are detailed in \S 4.2.  The 
inset images are to scale, and were synthesized by uniformly weighting the 
visibilities to highlight the structure at higher angular resolution.  
\label{results_trans}}
\end{figure}

\paragraph{Elias 24 --} This heavily-reddened classical T Tauri star is one of 
the brightest millimeter continuum sources in Ophiuchus.  In fact, it was 
initially excluded from the sample out of concern that the bright emission 
signaled contamination from an extended envelope.  However, there is no 
evidence for the spatial filtering of such large-scale emission; the SMA flux 
densities at both 880\,$\mu$m and 1.3\,mm are in excellent agreement with lower 
resolution single-dish photometry \citep{aw07,aw07b,andre94}.  Moreover, the 
central star is optically visible \citep{wilking05} and the infrared SED shape 
is incompatible with a substantial envelope 
\citep{bontemps01,evans03,barsony05}.  The 880\,$\mu$m continuum data presented 
here features a bright central core and a fainter, extended, and apparently 
asymmetric emission halo on larger scales.  Although the best-fit model for 
this source is able to reproduce the SED and visibilities rather well, any 
axisymmetric model will underpredict the extended emission to the east of the 
disk center at the $\sim$3$\sigma$ level (see Figure \ref{results_new}).  The 
origins of that extension remain unclear.

\begin{figure}
\epsscale{1.0}
\plotone{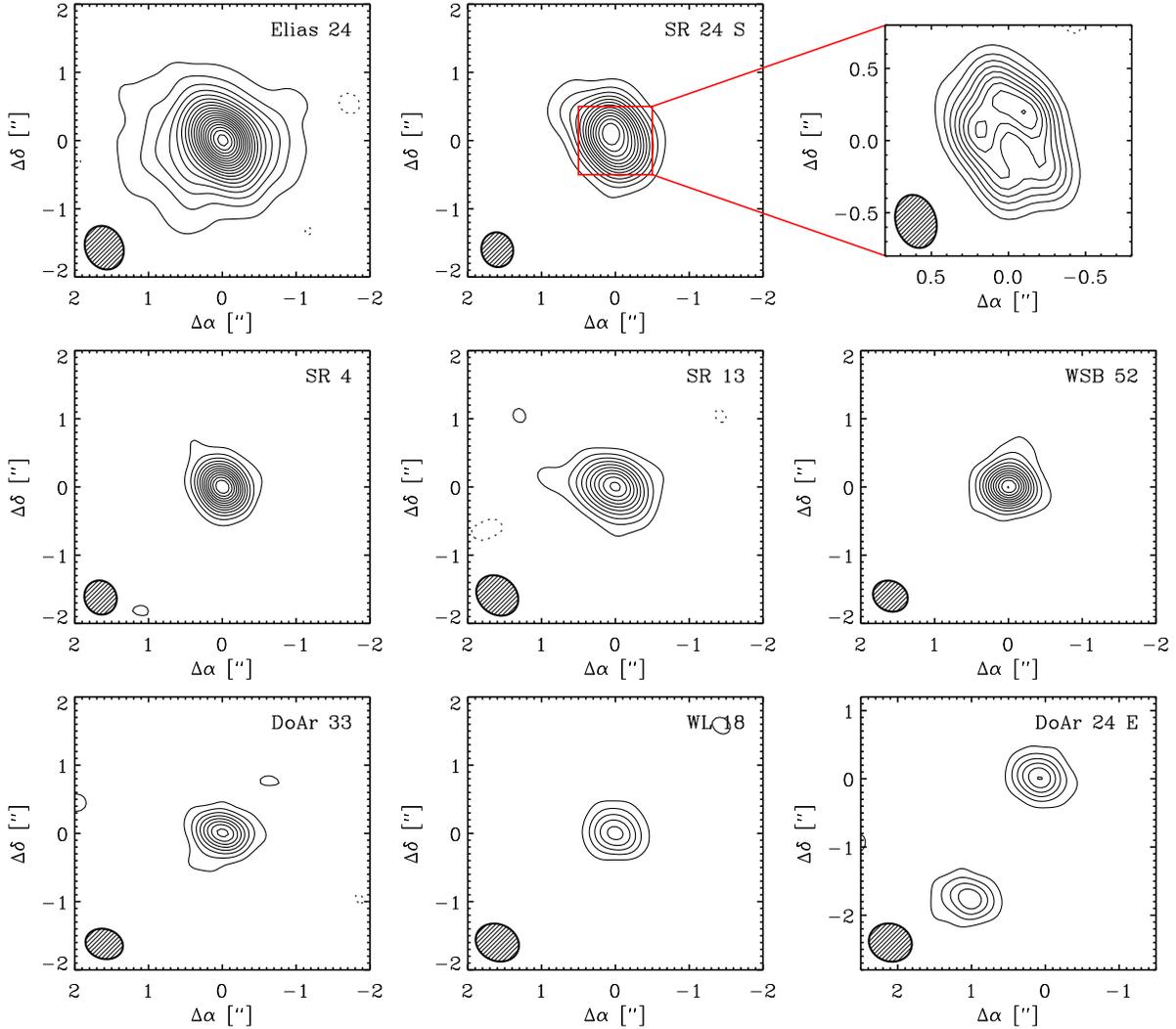}
\figcaption{Aperture synthesis images of the 880\,$\mu$m continuum emission 
from the 8 disk targets in the expanded sample.  Each panel is 4\arcsec\ 
(500\,AU) on a side.  Contours start at 3\,$\sigma$ and increase in 3\,$\sigma$ 
intervales (5\,$\sigma$ for Elias 24 and SR 24 S only; rms values are given in 
Table \ref{data_table}).  The synthesized beams are shown in the lower left 
corner of each panel.  Note the detection of roughly equal emission levels 
around both the optically-visible primary star DoAr 24 E and its infrared 
companion to the southeast, as well as the asymmetric ring-like emission 
morphology for the disk around SR 24 S (shown separately in detail).  
\label{images}}
\end{figure}

\paragraph{SR 24 S --} The hierarchical triple system SR 24 is composed of this 
K2 star and a close binary pair \citep[$0\farcs2 \approx 25$\,AU;][]{simon95} 
located $\sim$5\arcsec\ (625\,AU) to the north \citep{reipurth93}.  Both SR 24 
S and the SR 24 N binary exhibit excess emission from warm dust disks, bright 
H$\alpha$ lines indicative of substantial accretion flows, and extended 
emission from CO low-energy rotational transitions.  However, all of the 
continuum emission at millimeter wavelengths is produced by SR 24 S 
\citep{aw05a,patience08,isella09}.  The high angular resolution inset image of 
the SR 24 S disk in Figure \ref{images} reveals a resolved central emission 
cavity with an apparent brightness enhancement to the northeast.  The origins 
of this ring-like emission morphology are unclear, but could perhaps be 
generated by abrupt emissivity variations due to particle growth, a dramatic 
dissipation process driven by high-energy radiation from the star, or even 
tidal interactions with companion objects \citep[in this last case, 
see][]{mayama10}.    

Regardless of the cause, the model described in \S 3 was adjusted to account 
for the observed disk morphology.  As detailed in Paper I for similar cases, we 
adopted the simple modification of artificially decreasing the surface 
densities inside a radius $R_{\rm cav}$ by a factor $\delta_{\rm cav}$ and then 
attempted to fit the SMA visibilities and the component-resolved optical 
\citep{wilking05}, infrared \citep{jensen97,cutri03,evans03,mccabe06}, and 
radio photometry.  However, much like the case of the DoAr 44 disk described in 
Paper I, it is difficult to simultaneously account for the significant infrared 
excess and the lack of millimeter continuum emission near the SR 24 S stellar 
position.  Following the approach of \citet{espaillat07} for other 
``pre-transitional" disks, we proceeded by first finding a good match to the 
millimeter visibilities, and then artificially increasing the surface densities 
near the star until sufficient infrared emission was produced to match the SED 
(without affecting the millimeter emission).  A satisfactory result was 
achieved with a surface density profile scaled down by $\delta_{\rm cav} 
\approx 0.05$ inside $R = 2$\,AU, a factor of 20 lower than for a continuous 
disk, but $\sim$500$\times$ higher than for the region between 2\,AU and 
$R_{\rm cav} \approx 32$\,AU.  Obviously these adjustments are artificial and 
non-unique; a modeling effort focused on a more robust exploration of the 
detailed inner disk structures will be treated elsewhere.

\paragraph{SR 4 --} This star harbors a disk with a compact, centrally-peaked 
millimeter emission morphology.  The low foreground extinction allows for a 
particularly good characterization of the stellar luminosity from optical 
photometry, despite the low-level variability at those wavelengths 
\citep{herbst94}.  The infrared SED shows a relatively small excess at short 
wavelengths before a pronounced flattening from $\sim$8-24\,$\mu$m 
\citep{cutri03,evans03}.  Unfortunately there is little additional information 
in the far-infrared to better constrain the SED morphology, as the {\it 
Spitzer} 70\,$\mu$m images are contaminated by local nebulosity 
\citep{padgett08}.  This lack of data between the mid-infrared and millimeter 
regions of the SED results in only a relatively crude constraint on the 
vertical structure parameters (and thus temperatures) in this case.  

\paragraph{SR 13 --} Another hierarchical triple system 
\citep[see][]{schaefer06}, SR 13 includes a close binary pair \citep[13\,mas 
$\approx 1.6$\,AU;][]{simon95} and a tertiary companion located $\sim$0\farcs4 
(50\,AU) due east \citep{ghez93}.  Lacking sufficient component-resolved 
photometry for this system, we made a preliminary model ignoring the tertiary 
and assuming the primary is a single star.  The SED was constructed from the 
composite optical measurements of \citet{herbst94}, infrared data from 2MASS 
and the {\it Spitzer} c2d survey \citep{cutri03,evans03}, and single-dish radio 
observations \citep{andre94,aw07b}.  Despite this simplification, the SMA data 
can potentially resolve another disk around the tertiary.  There is a modest 
eastern extension in the emission map, and a potential null near 
400\,k$\lambda$ in the deprojected visibility profile (Figure 
\ref{results_new}).  These features could be produced either by a cavity in a 
circum-system disk or by a marginally resolved additional disk around the 
spatially offset tertiary.  With the available sensitivity, it is difficult to 
differentiate these scenarios.  Needless to say, the model disk structure 
inferred in this case should be treated with appropriate caution.

\paragraph{WSB 52 --} This classical T Tauri star with spectral type M1 
exhibits a compact millimeter continuum emission distribution typical of the 
fainter disks in the sample.  The SED used in the model fits was constructed 
from the red-optical photometry of \citet{wilking05}, infrared measurements 
from 2MASS \citep{cutri03}, {\it Spitzer} c2d \citep{evans03}, and 
\citet{padgett08}, and a 1.3\,mm single-dish flux measurement from 
\citet{stanke06}.

\paragraph{DoAr 33 --} The infrared excess emission around this K4 star is 
remarkably faint compared to the typical young disk.  We constructed a SED from 
red-optical photometry \citep{wilking05}, the 2MASS and {\it Spitzer} c2d 
programs \citep{cutri03,evans03}, and single-dish radio measurements 
\citep{andre94,aw07b}.  Only a very weak continuum excess is present shortward 
of $\sim$8\,$\mu$m.  At longer wavelengths where the dust excess is brighter, 
the spectrum is very steep and blue.  Reproducing this infrared SED shape with 
our models required the use of a very flat vertical distribution of dust, 
resulting in comparatively cold midplane temperatures.  \citet{cieza10} argue 
that these infrared colors are indicative of significant dust evolution.  The 
flat structure inferred here could be interpreted in that context as evidence 
for advanced dust sedimenation to the midplane, or perhaps the sign of 
extensive shadowing of the outer disk due to some perturbed inner disk 
structure.  In any case, the coupling of the rare infrared SED, bright 
millimeter emission, and weak signatures of accretion \citep{cieza10} suggest 
that this disk is worthy of further scrutiny.

\paragraph{WL 18 --} This star harbors the faintest disk in the sample, due in 
part to its small size ($R_c = 14$\,AU).  Its visibility profile shows that the 
millimeter emission is only marginally resolved on the longest SMA baselines.  
The SED compiled here is sparse due to high foreground extinction, with 
infrared data from 2MASS and the {\it Spitzer} c2d project 
\citep{cutri03,evans03}, and a single-dish 1.3\,mm flux density measured by 
\citet{motte98}.  That same extinction makes an estimate of the underlying 
stellar luminosity difficult, which in turn contributes an uncertainty to the 
thermal structure of the disk.  The apparently low luminosity of the star 
relative to others with the same spectral type (K7) leads to an uncomfortably 
large age estimate (11\,Myr).  

\begin{figure}
\epsscale{0.6}
\plotone{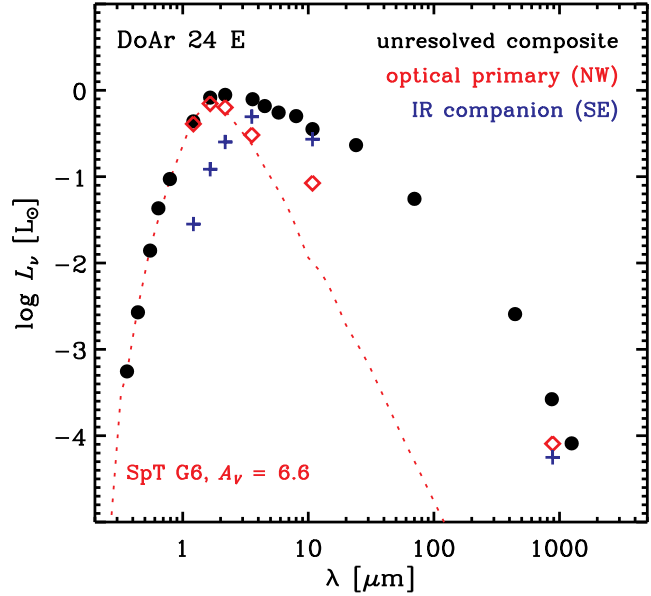}
\figcaption{The SED for the combined DoAr 24 E system ({\it black circles}) is 
shown together with the component-resolved SEDs of the optical primary ({\it 
red}) and infrared companion ({\it blue}) that lie to the northwest and 
southeast in Figure \ref{images}, respectively.  A model stellar photosphere 
for the optical primary is shown as a dashed red curve.  Unlike those shown in 
Figure \ref{results_new}, these SEDs have not been corrected for extinction due 
to the uncertainty in the nature of the infrared companion source.  
\label{doar24e}}
\end{figure}

\paragraph{DoAr 24 E --} This spectral type G6 weak-lined T Tauri star has a 
companion roughly 2\arcsec\ to the southeast \citep{ghez93,reipurth93} that 
becomes substantially brighter than the primary at wavelengths longer than 
$\sim$3\,$\mu$m \citep[e.g.,][]{mccabe06}.  The unresolved composite SED has 
been attributed to the photosphere of the primary and dust around this infrared 
companion.  However, resolved measurements at near- and mid-infrared 
wavelengths \citep{prato03,barsony05,mccabe06} indicate that the primary does 
show some very weak infrared excess at least at $\sim$10\,$\mu$m.  Regardless, 
most of the unresolved millimeter emission, and therefore the circumstellar 
dust mass, was assumed to be generated by the infrared companion 
\citep{motte98,aw07b}.  The resolved 880\,$\mu$m image of this system in Figure 
\ref{images} clearly demonstrates that this is not the case: each stellar 
component hosts a dust disk with roughly equal amounts of millimeter emission.  
Neither individual disk is clearly resolved.  The composite and 
component-resolved SEDs for this system are shown together in Figure 
\ref{doar24e}, along with a potential model for an extincted stellar 
photosphere from the primary.  Since very little resolved information is 
available and the nature of the infrared companion is uncertain, we have not 
modeled the circumstellar material in this system.  There is potentially great 
interest in doing so with some modifications to the standard technique 
presented here.  It would be worthwhile to better understand the lack of 
accretion onto the primary despite the presence of a substantial dust mass, as 
well as the potential origin of the very red SED of the companion object.

\section{Discussion}

\subsection{Constraints on Viscous Evolution}

We have used a radiative transfer modeling technique to extract the dust 
density structures for a significant sample of protoplanetary disks based on 
high angular resolution observations of their millimeter continuum emission.  
Those data and the SEDs for each source were reproduced well using a simple 
model for the two-dimensional density structure of a viscous accretion disk, 
with a parametric surface density profile that varies with radius like a 
power-law with an exponential taper at large radii (see Eq.~[1]).  This 
specific form of $\Sigma$ corresponds to the \citet{lyndenbell74} similarity 
solutions for the evolution of a thin accretion disk in Keplerian rotation 
around a (stellar) point mass, with a viscosity $\nu \propto R^{\gamma}$ that 
does not vary with time \citep[see also][]{hartmann98}.  This type of model - 
without a sharply truncated outer edge - is favored observationally, as it 
provides a natural explanation for the optical absorption profiles of 
silhouette disks in Orion \citep{mccaughrean96} and reconciles the apparent 
discrepancies in the observed spatial extents of the dust and CO line emission 
for nearby resolved disks \citep{hughes08}.  

In the context of these models, Figure \ref{histograms}b demonstrates that 
there is a relatively narrow distribution of the parameter $\gamma$ that 
describes the spatial distribution of mass in the disk, consistent with a 
median $\langle \gamma \rangle = 0.9\pm0.2$ when the sample is considered 
together (estimates for individual disks range from $\gamma = 0.4$-1.1).  These 
values fall at the high end of the wider distribution of $\Sigma$ gradients 
inferred by \citet[][where $\gamma$ ranges from -0.8 to 0.8]{isella09} using a 
1.3\,mm continuum survey of Taurus disks with slightly poorer angular 
resolution (0.7-1.0\arcsec), and therefore probing the mass at larger disk 
radii.  Although there is substantial overlap in the $\gamma$ distributions 
from both surveys (particularly when the disks with large central cavities are 
ignored), the differences can be attributed to the distinct approaches for 
interpreting the data and there is no obvious means of reconciliation at 
present.  Focusing on our results, we infer that the mass in $\sim$1\,Myr-old 
disks has a similar radial distribution regardless of the wide range of total 
masses ($M_d \approx 0.001$-0.1\,M$_{\odot}$) probed in this sample.  This 
implies that whatever mechanism is responsible for generating the viscosity in 
these disks insures that it has a roughly linear dependence on radius ($\gamma 
\approx 1$).  

\begin{figure}
\epsscale{0.6}
\plotone{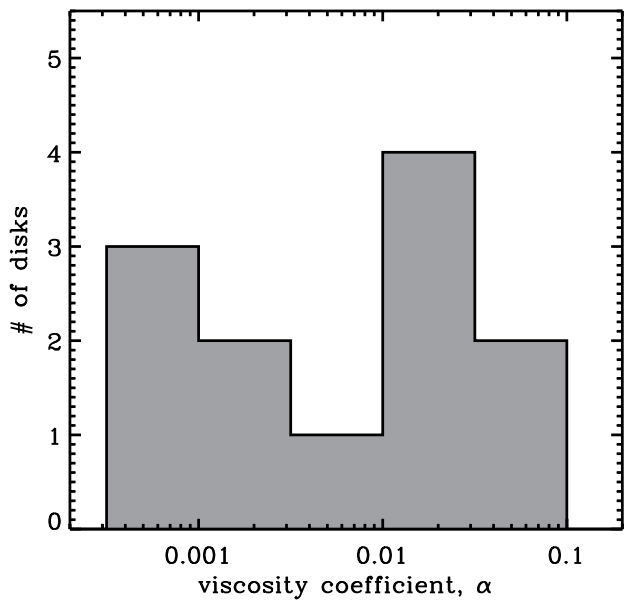}
\figcaption{The distribution of viscosity coefficients, $\alpha$, inferred for 
this sample from the disk structures and mass accretion rates.  The individual 
$\alpha$ values listed in Table \ref{viscous_table} are still quite uncertain, 
although the broad range is commensurate with simulations where the disk 
viscosities are sustained by MHD turbulence.  Given the small sample size, 
$\alpha$ uncertainties, and potential $\dot{M}_{\ast}$ bias, the bimodal 
appearance of the distribution is not significant. \label{vischist}}
\end{figure}

Since the physical origin of that viscosity is unclear, we adopt a simple 
prescription $\nu = \alpha c_s H$, where $c_s$ is the sound speed and the 
viscosity coefficient $\alpha$ describes the efficiency of the angular momentum 
transport \citep{shakura73}.  Using the density structures and measurements of 
mass accretion rates $\dot{M}_{\ast}$ (see Table \ref{viscous_table} for 
references), we can estimate the value of the viscosity coefficient $\alpha(R) 
\approx \dot{M}_{\ast} (R/R_c)^{\gamma}/3 \pi \Sigma_c c_s H$ (see the Appendix 
in Paper I for details).  The resulting distribution of $\alpha$ at $R = 
10$\,AU is shown in Figure \ref{vischist}, with values for individual disks 
listed in Table \ref{viscous_table}.  Note that for the typical midplane 
temperature distribution $T \propto R^{-q}$, the viscosity coefficient is 
proportional to $R^{\gamma}/c_s H \sim R^z$, where $z \approx \gamma+q/2-\psi - 1$.  The model parameters and temperature distributions inferred for this 
sample have $z \approx 0.0\pm0.3$, meaning $\alpha$ does not vary by more than 
a factor of $\sim$2-4 across the disk.  The inferred distribution of these 
viscosity coefficients appears bimodal, with peaks just below $\alpha \approx 
0.001$ and above 0.01.  However, given the small sample, those peaks are not 
statistically significant; they simply represent the clustering of the 
accretion rates in this sample.  The range of inferred $\alpha$ values is in 
reasonable quantitative agreement with magnetohydrodynamics simulations where 
the viscosity is generated by turbulence from the magnetorotational instability 
in slightly ionized disks \citep{hawley95,stone96,fleming03,fromang07}.  
However, we should caution that these $\alpha$ values are not yet well 
constrained: they suffer the combined uncertainties in the accretion rates and 
disk structure parameters, as well as from systematic issues with the inherent 
assumptions in their derivation (e.g., the dust traces the gas, viscous heating 
is insignificant, etc.).  They represent only a first exploration of an 
improved empirical understanding of disk viscosities.  

The interactions of gravitational and viscous torques control the evolution of 
disk structure over the vast majority of the disk lifetime.  This viscous 
evolution process has two important, observable effects on that structure.  
First, the coupling of the viscosity and the Keplerian orbital shear drives a 
net mass flow toward small radii, where material can be magnetically channeled 
onto the central star.  That mass flow produces bright H emission lines, and 
the shock generated when it impacts the stellar surface gives rise to a strong 
ultraviolet continuum; both tracers can be used to estimate $\dot{M}_{\ast}$ 
\citep{muzerolle98a,muzerolle98b,gullbring98,gullbring00}.  To compensate for 
the angular momentum dissipated in that process, the material in the outer disk 
(beyond some radius $R_t$; see Table \ref{viscous_table} and Paper I) is spread 
to larger radii.  The results of that viscous diffusion - decreased average 
densities (i.e., $M_d$) and increased sizes ($R_c$) - can be probed with 
resolved observations of the millimeter continuum emission.  In principle, 
tracking these observational signatures as a function of time could provide 
strong constraints on the viscous evolution process.  \citet{isella09} claim a 
significant correlation between their parameterization of $R_c$ and stellar age 
in a sample of Taurus disks.  No such correlation is evident in our larger 
sample of Ophiuchus disks.  More importantly, the search for such a trend 
within a single star-forming region is probably premature; the ages of 
individual young stars can not be determined with sufficient accuracy to infer 
such evolutionary behavior \citep[e.g.,][and references 
therein]{hillenbrand09}, even though it may exist.  Rather, the inferred disk 
structures could be considered representative of the diversity of evolutionary 
states and/or initial conditions at a ``snapshot" in the evolution sequence 
corresponding to some median cluster age (in this case, $\sim$1\,Myr).  

\begin{figure}
\epsscale{0.6}
\plotone{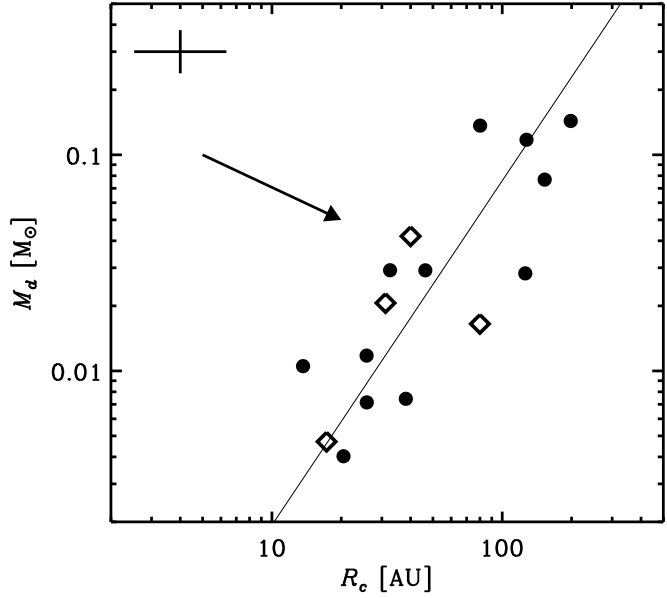}
\figcaption{A 3.3\,$\sigma$ correlation between the disk masses and 
characteristic radii.  The solid line shows a power-law that provides a 
reasonable match to this relationship, where $M_d \propto R_c^x$ with $x = 
1.6\pm0.3$.  Diamond symbols mark the disks with large central emission 
cavities, and the cross to the upper left is representative of the typical 
parameter uncertainties.  The arrow denotes the generic direction of evolution 
implied by the conservation of angular momentum.  The range along the 
correlation is representative of the range of angular momenta incorporated into 
the disks when they formed, but the shape of the correlation encodes both these 
initial conditions and potentially a range of viscous timescales. \label{M_R}}
\end{figure}

While there may not be much difference in the shapes of the $\Sigma$ profiles 
($\gamma$), this sample includes disks with a wide variety of masses ($M_d$) 
and characteristic size scales ($R_c$).  As was demonstrated in Figure 
\ref{median_vis}, the disks with brighter millimeter emission have 
systematically larger masses {\it and sizes}.  This relationship is illustrated 
more directly in Figure \ref{M_R}, with a significant correlation 
(3.3\,$\sigma$; Spearman rank coefficient of 0.85) between $R_c$ and $M_d$ that 
can be approximated as a power-law, $M_d \propto R_c^{1.6 \pm 0.3}$.  The 
physical origin of the correlation is not clear.  One potential explanation is 
that it reflects the range of initial conditions and viscous properties 
inherited at the disk formation epoch.  In that case, note that the sense of 
the correlation is almost perpendicular to the evolutionary paths for 
individual disks that conserve angular momentum, $M_d \propto R_c^{-1/2}$.  The 
spread along the correlation is then representative of the distribution of 
angular momenta imparted by the collapse of the parent molecular cloud cores.  
Following \citet{isella09}, we can estimate the specific angular momenta in 
these cores that are needed to reproduce the inferred disk structures.  
Assuming the centrifugal radius in the disk corresponds to the radius that 
contained $\sim$90\%\ of the disk mass at the formation epoch (here defined as 
approximately twice an initial scaling radius, $\sim$2$R_1$), then the specific 
angular momentum can be written
\begin{equation}
j \approx 2 \times 10^{20} \left(\frac{R_1}{25\,\rm{AU}}\right)^{0.5}\left(\frac{M_{\ast}}{{\rm M}_{\odot}}\right)^{-1.5}\left(\frac{R_{core}}{0.1\,{\rm pc}}\right)^2 {\rm cm}^2 \,\,\, {\rm s}^{-1},
\end{equation}
where $R_{core}$ is the core radius \citep[see][]{hueso05}.  Precise values for 
the initial scaling radius are uncertain (see below), but we can estimate upper 
limits on the $j$ values since $R_1 \le R_c$ by definition (see Paper I).  
Substituting $R_1 = R_c$, $R_{core} = 0.1$\,pc, and the stellar masses in Table 
\ref{stars_table} into Eq.~(2), we find $\log{j} \approx 19.7$-20.9 in units of 
cm$^2$\,s$^{-1}$.  Radio observations of rotating molecular cloud cores have 
been used to infer similar or higher values, $\log{j} \approx 19.6$-22.2 
\citep{goodman93,barranco98,caselli02}.  However, recent numerical simulations 
have demonstrated that those measurements tend to overestimate $j$ by roughly 
an order of magnitude \citep{dib10}.  If that is the case, then the range of 
disk angular momenta inferred here are actually in good quantitative agreement 
with those determined for analogs of their parental molecular cloud cores.  The 
low end of the $j$ distribution postulated by \citet{dib10} and not probed here 
could then be explained by a selection effect (perhaps our flux-limited sample 
only recovers disks with large $j$ values), an unspecified mechanism for 
angular momentum loss (e.g., outflows or magnetic braking), and/or that $R_1 
\ll R_c$ in at least some cases.

Unfortunately, an unambiguous explanation of the {\it shape} of the $M_d$-$R_c$ 
correlation in this scenario is not possible, due to the fundamental degeneracy 
between the initial disk structures and the rates at which they evolve.  The 
values of $M_d$ and $R_c$ change with time such that $M_d = M_{d,0}/{\cal 
T}^{1/2(2-\gamma)}$ and $R_c = R_1 {\cal T}^{1/(2-\gamma)}$, where $M_{d,0}$ 
and $R_1$ were the initial disk mass and characteristic radius, respectively, 
and ${\cal T}$ is a dimensionless parameter that tracks how many viscous 
timescales ($t_s$) have elapsed, ${\cal T} = 1 + t/t_s$ \citep[see Paper 
I;][]{hartmann98}.  That evolution is described by 2 equations with 3 unknown 
parameters, \{$M_{d,0}$, $R_1$, $t_s$\}: the evolution rate can not be uniquely 
disentangled from the initial conditions.  However, the lack of significant 
outliers to the correlation may be related to the viscous evolution process 
itself.  The ``small and massive" region in the upper left of Figure \ref{M_R} 
is depopulated because these disks spend only a short time with such compact 
density configurations early in their evolution ($t \ll 1$\,Myr).  Likewise, 
the ``large and low-mass" region to the lower right in this parameter-space is 
empty because these $\sim$1\,Myr-old disks simply have not had enough time to 
evolve into it.  

Alternatively, the inferred $M_d$-$R_c$ correlation could be the result of 
coupling the viscous evolution scenario described above with an internal 
dissipation process.  Recent theoretical calculations have suggested that the 
high-energy (particularly far-ultraviolet, or FUV) irradiation of the disk 
surface by the central star can drive significant mass loss from large disk 
radii ($R > 30$\,AU) in a photoevaporative flow \citep{gorti09a,gorti09b}.  In 
this scenario, the smaller, low-mass disks in this sample should be associated 
with an advanced stage in this photoevaporation process, perhaps because their 
central stars have more intense FUV radiation fields.  Unfortunately, direct 
constraints on the FUV emission from these extincted sources in the Ophiuchus 
clouds are rare.  Instead, we can approximate the FUV luminosity as the sum of 
a fraction of the accretion luminosity ($\propto M_{\ast} \dot{M}_{\ast} / 
R_{\ast}$) and a chromospheric component \citep[$\approx 5 \times 10^{-4} 
L_{\ast}$; see][]{gorti09b}.  However, we find no evidence for an expected 
anti-correlation between disk masses/sizes and the FUV (or X-ray) luminosities 
in this sample: if anything, the larger and more massive disks tend to have 
more intense FUV radiation fields, whereas the less massive, smaller disks show 
a wide range of FUV luminosities.  While this certainly does not exclude a role 
for photoevaporation in shaping disk structures, it does suggest that diverse 
initial conditions and viscous timescales are more significant contributors to 
the observed $M_d$-$R_c$ correlation at an age of $\sim$1\,Myr.

Comparisons of disk properties in different stellar populations will be 
required to definitively assess the relative impacts that initial conditions, 
viscous evolution, and dissipation effects like photoevaporation have on 
shaping disk density distributions over time.  The key to disentangling the 
initial conditions from the viscous evolution timescale is to search for and 
characterize structural correlations, like the $M_d$-$R_c$ relationship found 
here, in disks around stars with a range of ages.  In a scenario where viscous 
evolution dominates, the disks embedded in the envelopes of Class 0/I sources 
should be massive and compact, providing direct constraints on the 
distributions of initial masses ($M_{d,0}$) and sizes ($R_1$).  Conversely, the 
disks in older star-forming regions are expected to be larger and less massive, 
unless or until FUV photoevaporation can effectively dissipate the mass 
reservoir in the outer disk.  Tracking the evolution of disk masses and sizes 
in these older stellar populations can provide crucial insights into the 
distribution of viscous timescales, as well as into the timeframe where 
photoevaporation may dissipate much of the disk mass reservoir.  To probe disk 
structures in these older star clusters that are typically at larger distances, 
instruments with increased sensitivity to millimeter continuum emission and 
access to higher angular resolution will be required.  Fortunately, there is 
great promise for a breakthrough in our understanding of viscous evolution and 
disk dissipation in the near future, based on extensive millimeter continuum 
surveys of disk populations with the Atacama Large Millimeter Array (ALMA).

\subsection{Implications for Planet Formation}

Ultimately, the drive to understand this viscous evolution process lies in the 
desire to constrain how mass is re-distributed over time, and therefore where 
and when the conditions in a given disk are suitable for planet formation.  
All planet formation models require that the disk exceeds some density 
threshold as a necessary (but perhaps not sufficient) condition for making a 
planet.  In light of the new constraints on disk densities presented here and 
in Paper I, it is natural to compare them with the two basic recipes for giant 
planet formation; disk instability \citep[e.g.,][]{boss97} and core accretion 
\citep[e.g.,][]{pollack96}.  In the disk instability model, an over-dense 
region with a sufficiently short cooling time can fragment out of the global 
disk structure and precipitate a bound gaseous protoplanet very rapidly 
\citep[$\sim$10$^{3-4}$\,yr; see the recent review by][]{durisen07}.  Typically 
those conditions are only met at large disk radii \citep{boley06}; such 
protoplanets must then migrate inward to reproduce the shorter period orbits 
observed in the Solar System and elsewhere.  In the core accretion model, a 
giant planet is produced from a relatively slow ($\sim$1-10\,Myr) collisional 
growth process that first builds up a large solid core and then rapidly 
accretes a massive gaseous envelope \citep[e.g.,][]{hubickyj05,alibert05}.  

Assuming the dust traces 1\%\ of the gas mass, the disk structures presented 
here are stable against gravitational fragmentation, with the minimum local 
Toomre $Q$ values ranging from $\sim$5-50 in most cases \citep[$Q = c_s \Omega 
/ \pi G \Sigma$, where $\Omega$ is the Keplerian orbital 
velocity;][]{toomre64}.  The large, massive disks around DoAr 25 and GSS 39 are 
potentially exceptions, with $Q$ values approaching 2 at radii of $\sim$65 and 
150\,AU, respectively.  However, we should caution that the disk structures in 
both cases are particularly uncertain (see \S 4.2 of Paper I).  While none of 
the sample disks are clear candidates for the efficient operation of the disk 
instability mechanism for planet formation in the current epoch (i.e., $Q \le 
1.7$), their structures may have been less stable at earlier times.  Even 
without invoking some kind of mass loading mechanism \citep{boley09}, viscous 
evolution implies that these disks originally had denser, more compact 
structures, which might suggest a formerly lower minimum $Q$-value and/or 
radius where the disk is least stable to fragmentation.  A quantitative 
exploration of the history of $Q(R)$ in these disks grounded in our constraints 
on their viscous properties may well be worthwhile, but would require new 
radiative transfer calculations that incorporate some model for the thermal 
evolution of the disk material in each case.

A similar comparison with the core accretion model is more challenging, 
primarily because resolution limitations do not yet permit a direct observation 
of the disk material inside $R \approx 20$\,AU.  Typically, core accretion 
model calculations impose a scaled Minimum Mass Solar Nebula (MMSN) density 
structure for an initial disk, where $\Sigma \propto R^{-3/2}$ 
\citep{weidenschilling77} and a total mass set to produce a formation 
efficiency commensurate with the observational constraints on disk lifetimes.  
Although the MMSN surface density profile is too steep compared to those 
derived here (where $\Sigma \propto R^{-1}$ at the relevant radii), Figure 
\ref{sigma} demonstrates that the surface densities for the sample are 
generally compatible with those expected in the 20-40\,AU region of the 
primordial Solar disk.  The shape of the $\Sigma$ profile should not adversely 
affect the likelihood of planet formation if sufficient mass is available, but 
it is expected to have an impact on the orbital architecture and migration 
properties of any resulting planetary system 
\citep{kokubo02,chambers02,raymond05,crida09}.  Some core accretion 
simulations have started to employ viscous disk density structures similar to 
those derived here \citep[e.g.,][]{alibert05,hueso05}, but they still must 
populate that density structure with solid bodies orders of magnitude larger 
than those responsible for the observed millimeter emission.

This last point highlights an important uncertainty in deriving densities from 
observations of young circumstellar disks: the conversion of emission to mass 
does not account for particles much larger than the observing wavelength.  
Since this potential for substantially under-estimating the disk densities was 
already elaborated in Paper I, we will not dwell on it here.  The problem is 
not likely to be severe in the regions that can currently be probed 
observationally ($R \ge 20$\,AU), as even optimistic estimates of particle 
growth timescales at those radii do not predict a substantial population of 
large solids \citep[e.g.,][]{dullemond05}.  However, at smaller disk radii 
those growth timescales can be considerably shorter, generating a dust 
emissivity that varies with radius inside the planet formation zone.  More 
generally, it is worthwhile to emphasize that the observations are not yet 
sensitive to the midplane material for $R < 20$\,AU.  Inside that resolution 
limit, the densities could be dramatically different than inferred here if, for 
example, the dust size distribution is changed or the effective viscosity 
profile is modified \citep[e.g.,][]{zhu10}.  New data from ALMA and the 
Expanded Very Large Array (EVLA) will help alleviate these uncertainties at 
small radii, and should also help empirically guide our understanding of disk 
viscosities.  

But despite our general ignorance of the inner disk structures, a subset of 
young disks show unambiguous evidence for evolution in the planet formation 
zone.  In this SMA survey of 17 disks, selected primarily for their millimeter 
luminosities, 4 of them - around SR 24 S, SR 21, WSB 60, and DoAr 44 - have 
large central cavities with significantly diminished dust emission.  The 
characteristic ring-like emission morphologies and distinctive visibility nulls 
for these disks indicate that the radial transition in the millimeter optical 
depth across the cavity edge is rather sharp, commensurate with our simple 
models that employ a density contrast of $\sim$10$^2$-$10^4$ within a small 
radial range.  In 3 of these 4 disks, a small amount of micron-sized dust 
grains must reside inside the cavity to account for their observed infrared 
excess emission.  Outside of their central cavities, these disks have $\Sigma$ 
profiles comparable to the disks with continuous dust distributions.  At this 
point, the physical mechanism responsible for the dust evolution in these disk 
cavities remains a subject of active debate \citep[e.g., 
see][]{dalessio05,najita07}.  They could be produced by an abrupt emissivity 
decrease due to grain growth in a localized region of low turbulence 
\citep[e.g., an MRI-inactive ``dead" zone;][]{ciesla07,zhu10}.  Alternatively, 
they may represent a true lack of material in the inner disk, cleared out by a 
photoevaporative wind and viscous draining \citep{alexander07} or tidal 
interactions with faint companions - perhaps even young planetary systems 
\citep[e.g.,][]{lubow06}.

\section{Summary}

We have conducted a high angular resolution (down to FWHM scales of 0\farcs3 
$\approx 40$\,AU) Submillimeter Array survey of the 880\,$\mu$m continuum 
emission from 17 protoplanetary disks in the $\sim$1\,Myr-old Ophiuchus 
star-forming region \citep[see also][]{andrews09}.  Using a two-dimensional 
parametric model for the structure of a viscous accretion disk and a Monte 
Carlo radiative transfer code to simultaneously reproduce the millimeter 
visibilities and broadband SEDs, we have measured the spatial distribution of 
mass in these disks.  The key conclusions from this survey include:
\begin{enumerate}
\item Assuming the viscosity scales with radius like a power-law ($\nu \propto 
R^{\gamma}$) and the surface densities like $\Sigma \propto (R/R_c)^{-\gamma} 
\exp{[-(R/R_c)^{2-\gamma}]}$, the disks in this survey exhibit a relatively 
narrow range of density/viscosity gradients, $\gamma = 0.4$-1.1.  Taken 
together as a sample, these results are consistent with a single (median) 
value, $\langle \gamma \rangle = 0.9 \pm 0.2$, independent of the millimeter 
luminosity (disk mass) or the stellar properties.
\item The disk masses ($M_d$) and characteristic radii ($R_c$) are correlated: 
massive disks tend to be larger (i.e., brighter disks are more resolved).  The 
3.3\,$\sigma$ correlation is described well by a power-law relation, $M_d 
\propto R_c^{1.6\pm0.3}$ (with a residual scatter of $\sim$0.3\,dex).  Rough 
estimates of the range of angular momenta in these disks are found to be 
comparable to the values inferred for rotating molecular cloud cores.  If it 
can be measured in disk samples with a well-established range of ages, the 
$M_d$-$R_c$ relationship can be used to deduce a mean viscous timescale and the 
range of initial conditions imparted by the disk formation process.  
\item Based on the derived physical conditions in these disks and literature 
measurements of their mass accretion rates, we have made crude estimates of 
their viscosity coefficients \citep[$\alpha$;][]{shakura73}.  The inferred 
$\alpha$ values range from 0.0005-0.08, with a median $\langle \alpha \rangle 
\approx 0.01$.  This range of values is commensurate with numerical simulations 
where the effective viscosities are generated by turbulence from the 
magnetorotational instability in slightly ionized disks.
\item The inferred disk surface densities in the $\sim$20-40\,AU range are in 
good quantitative agreement with those expected for the primordial disk that 
gave rise to the giant planets Uranus and Neptune (i.e., the outer part of the 
Minimum Mass Solar Nebula).  Current limitations on angular resolution prevent 
more direct constraints on $\Sigma$ in the inner disk ($R < 20$\,AU).  All of 
the sample disks appear to be gravitationally stable (typical minimum Toomre 
$Q$ values range from $\sim$5-50).  However, these results do not rule out 
periods where the disk instability mechanism for planet formation at earlier 
epochs ($\ll$1\,Myr). 
\item Regions of significantly diminished millimeter emission were resolved at 
the centers of 4/17 disks in the sample (SR 24 S, SR 21, DoAr 44, and WSB 60).  
Simple models for these ``transition" disks can reproduce the observations well 
if the densities are decreased by a factor of $\ge$100 inside a radius 
$R_{\rm cav} \approx 20$-40\,AU.  The disk properties outside these central 
cavities are comparable to the other disks in the sample.  The significant 
infrared excesses still present in 3/4 of these disks with resolved central 
cavities indicate that some small, warm, dust particles remain near their 
central stars.  The implied dust structures in those cases are commensurate 
with the opening of large gaps in the planet formation zones of these disks.  
\end{enumerate}

\acknowledgments We are grateful to an anonymous referee for suggestions that 
helped improve the clarity of this article.  The SMA is a joint project 
between the Smithsonian Astrophysical Observatory and the Academia Sinica 
Institute of Astronomy and Astrophysics and is funded by the Smithsonian 
Institution and the Academia Sinica.  Support for this work was provided by 
NASA through Hubble Fellowship grant HF-01203.01-A awarded by the Space 
Telescope Science Institute, which is operated by the Association of 
Universities for Research in Astronomy, Inc., for NASA, under contract NAS 
5-26555.  D.~J.~W.~acknowledges support from NASA Origins Grant NNG05GI81G.  
A.~M.~H.~acknowledges support from a National Science Foundation Graduate 
Research Fellowship.

\clearpage

\begin{deluxetable}{lcccll}
\tablecolumns{6}
\tablewidth{0pc}
\tablecaption{SMA Observing Journal\label{obs_journal}}
\tablehead{
\colhead{Name} & \colhead{$\alpha$ [J2000]} & \colhead{$\delta$ [J2000]} & \colhead{Array} & \colhead{UT Date} & \colhead{Alt.~Name} \\
\colhead{(1)} & \colhead{(2)} & \colhead{(3)} & \colhead{(4)} & \colhead{(5)} & \colhead{(6)}}
\startdata
Elias 24   & 16 26 24.08 & $-$24 16 13.7 & V & 2009 March 14 & WSB 31, YLW 32 \\
           &             &               & C & 2009 May 2    \\
SR 24 S    & 16 26 58.51 & $-$24 45 37.0 & V & 2009 March 25 & DoAr 29, Haro 1-7 \\
           &             &               & C & 2009 May 4    \\
SR 4       & 16 25 56.16 & $-$24 20 48.5 & V & 2009 March 12 & V2058 Oph, AS 206 \\
           &             &               & V & 2009 March 25 & \\
           &             &               & C & 2009 May 4    \\
SR 13      & 16 28 45.27 & $-$24 28 19.2 & V & 2009 March 12 & V853 Oph, HBC 266 \\
           &             &               & C & 2009 May 2    & \\
WSB 52     & 16 27 39.44 & $-$24 39 15.7 & V & 2009 March 2  & ROXs 27 \\
           &             &               & V & 2009 March 28 & \\
           &             &               & C & 2009 May 4    \\
DoAr 33    & 16 27 39.01 & $-$23 58 18.9 & V & 2009 March 2  & WSB 53, ROXs 30C \\
           &             &               & V & 2009 March 28 & \\
           &             &               & C & 2009 May 4    & \\
WL 18      & 16 26 48.98 & $-$24 38 25.4 & V & 2009 March 13 & GY 129 \\
           &             &               & C & 2009 May 2    \\
DoAr 24 E  & 16 26 23.37 & $-$24 20 59.8 & V & 2009 March 13 & Elias 22, GSS 31 \\
           &             &               & C & 2009 May 2  
\enddata
\tablecomments{Col.~(1): Disk name.  Cols.~(2) \& (3): Centroid coordinates, 
determined as described in the text (\S 2).  The coordinates listed for DoAr 24 
E correspond to the optically-visible component to the northwest.  Col.~(4): 
Array configuration; V = very extended (68-509\,m baselines) and C = compact 
(16-70\,m baselines).  Col.~(5): UT date of observation.  Col.~(6): Common 
alternative identifications.}
\end{deluxetable}

\clearpage

\begin{deluxetable}{lrcrcccc}
\tablecolumns{8}
\tablewidth{0pc}
\tablecaption{Continuum and CO Synthesized Map Properties\label{data_table}}
\tablehead{
\colhead{Disk} & \multicolumn{3}{c}{continuum} & \colhead{} & \multicolumn{3}{c}{CO $J$=3$-$2} \\
\cline{1-4} \cline{6-8} 
\colhead{} & \colhead{$F_{\nu}$} & \colhead{$\theta_b$} & \colhead{PA$_b$}  & \colhead{} & \colhead{rms} & \colhead{$\theta_b$} & \colhead{PA$_b$} \\
\colhead{} & \colhead{[mJy]} & \colhead{[\arcsec]} & \colhead{[\degr]} & \colhead{} & \colhead{[Jy]} & \colhead{[\arcsec]} & \colhead{[\degr]} \\
\colhead{(1)} & \colhead{(2)} & \colhead{(3)} & \colhead{(4)} & \colhead{} & \colhead{(5)} & \colhead{(6)} & \colhead{(7)}}
\startdata
Elias 24      & $890\pm3$ & $0.65\times0.51$ & 18 & & 0.19 & $2.13\times1.42$ & 51  \\
SR 24 S       & $545\pm3$ & $0.51\times0.43$\tablenotemark{a} & 89 & & 0.25 & $2.15\times1.44$ & 52  \\
SR 4          & $150\pm2$ & $0.50\times0.43$ & 15 & & 0.24 & $2.12\times1.45$ & 53  \\
SR 13         & $149\pm3$ & $0.63\times0.52$ & 39 & & 0.19 & $2.15\times1.41$ & 50  \\
WSB 52        & $147\pm3$ & $0.50\times0.43$ & 52 & & 0.25 & $2.16\times1.44$ & 52  \\
DoAr 33       & $ 80\pm2$ & $0.52\times0.43$ & 67 & & 0.25 & $2.17\times1.42$ & 53  \\
WL 18         & $ 51\pm3$ & $0.61\times0.52$ & 55 & & 0.19 & $2.19\times1.39$ & 50  \\
DoAr 24 E     & $ 49\pm2$\tablenotemark{b} & $0.60\times0.54$ & 58 & & 0.19 & $2.21\times1.37$ & 50  \\
\enddata
\tablecomments{Col.~(1): Disk name.  Col.~(2): Integrated continuum flux 
density and rms noise level per beam (does not include the $\sim$10\%\ flux 
calibration uncertainty).  Cols.~(3) \& (4): FWHM dimensions and position angle 
(measured east of north) of the synthesized beam for the continuum maps shown 
in Figure \ref{images}.  Col.~(5): The rms noise level per beam for an 
individual 0.70\,km s$^{-1}$ channel in the synthesized CO $J$=3$-$2 channel 
maps.  Cols.~(6) \& (7): FWHM dimensions and position angle of the synthesized 
beam for the channel maps.}
\tablenotetext{a}{The high-resolution inset map of the SR 24 S disk shown in 
Figure \ref{images} was generated with uniform visibility weighting and has a 
synthesized beam with dimensions $0\farcs37\times0\farcs26$ at PA = 14\degr.}
\tablenotetext{b}{The infrared companion source to the southeast is just 
slightly fainter, with an integrated flux density of $34\pm2$\,mJy.}
\end{deluxetable}

\begin{deluxetable}{lcccccccc}
\tablecolumns{10}
\tablewidth{0pt}
\tablecaption{Stellar Properties\label{stars_table}}
\tablehead{
\colhead{Name} & \colhead{SpT} & \colhead{$A_V$} & \colhead{$T_{{\rm eff}}$} & \colhead{$R_{\ast}$} & \colhead{$L_{\ast}$} & \colhead{$M_{\ast}$} & \colhead{age} & \colhead{ref} \\
\colhead{} & \colhead{} & \colhead{[mag]} & \colhead{[K]} & \colhead{[R$_{\odot}$]} & \colhead{[L$_{\odot}$]} & \colhead{[M$_{\odot}$]} & \colhead{[Myr]} & \colhead{} \\
\colhead{(1)} & \colhead{(2)} & \colhead{(3)} & \colhead{(4)} & \colhead{(5)} & \colhead{(6)} & \colhead{(7)} & \colhead{(8)} & \colhead{(9)}}
\startdata
{\it AS 205}  & K5 & 2.9  & 4250 & 3.7 & 4.0 & 1.0 & 0.5 & 1 \\
Elias 24      & K5 & 8.7  & 4250 & 4.2 & 5.1 & 1.0 & 0.4 & 2 \\
{\it GSS 39}  & M0 & 15   & 3850 & 2.3 & 1.0 & 0.6 & 1.0 & 3 \\
{\it AS 209}  & K5 & 0.9  & 4250 & 2.3 & 1.5 & 0.9 & 1.6 & 4 \\
{\it DoAr 25} & K5 & 2.7  & 4250 & 1.7 & 0.8 & 1.0 & 3.8 & 2 \\
SR 24 S       & K2 & 7.0  & 4990 & 2.8 & 4.4 & 2.0 & 2.4 & 3 \\
{\it WaOph 6} & K6 & 3.6  & 4205 & 3.2 & 2.9 & 0.9 & 0.7 & 5 \\
{\it SR 21}   & G3 & 6.3  & 5800 & 3.3 & 11  & 2.0 & 4.7 & 1 \\
{\it VSSG 1}  & M0 & 14   & 3850 & 3.1 & 1.9 & 0.6 & 0.7 & 6 \\
{\it WSB 60}  & M4 & 3.5  & 3370 & 1.3 & 0.2 & 0.3 & 3.0 & 2 \\
{\it DoAr 44} & K3 & 2.3  & 4730 & 1.7 & 1.3 & 1.4 & 7.1 & 7 \\
SR 4          & K7 & 1.3  & 4060 & 2.0 & 1.0 & 0.7 & 1.8 & 3 \\
SR 13         & M4 & 0.0  & 3370 & 1.9 & 0.4 & 0.3 & 1.5 & 2 \\
WSB 52        & M1 & 5.0  & 3750 & 1.9 & 0.6 & 0.5 & 1.5 & 3 \\
DoAr 33       & K4 & 3.7  & 4470 & 1.8 & 1.2 & 1.3 & 4.4 & 7 \\
WL 18         & K7 & 11   & 4060 & 1.1 & 0.3 & 0.8 & 11  & 2 \\
\enddata
\tablecomments{Col.~(1): Disk name (those in italics were modeled in Paper I).  
Col.~(2): Spectral type.  Col.~(3): Extinction.  Col.~(4): Effective 
temperature.  Col.~(5): Radius.  Col.~(6): Luminosity.  Col.~(7) and (8): Mass 
and age estimated from the \citet{siess00} pre-main-sequence models.  Col.~(9): 
Literature references for SpT and $A_V$: [1] - \citet{prato03}, [2] - 
\citet{wilking05}, [3] - \citet{luhman99}, [4] - \citet{herbig88}, [5] - 
\citet{eisner05}, [6] - \citet{natta06}, [7] - \citet{bouvier92}.}
\end{deluxetable}

\begin{deluxetable}{lccccc|ccc|cc}
\tablecolumns{11}
\tablewidth{0pt}
\tablecaption{Disk Structure Model Parameters: Continuous Cases\label{structure_table}}
\tablehead{
\colhead{Name} & \colhead{$M_d$} & \colhead{$\gamma$} & \colhead{$R_c$} & \colhead{$H_{100}$} & \colhead{$\psi$} & \colhead{$R_{{\rm in}}$} & \colhead{$i$} & \colhead{PA} & \colhead{$\tilde{\chi}^2_{\rm vis}$} & \colhead{$\tilde{\chi}^2_{\rm sed}$} \\
\colhead{} & \colhead{[M$_{\odot}$]} & \colhead{} & \colhead{[AU]} & \colhead{[AU]} & \colhead{} & \colhead{[AU]} & \colhead{[\degr]} & \colhead{[\degr]} & \colhead{} & \colhead{} \\
\colhead{(1)} & \colhead{(2)} & \colhead{(3)} & \colhead{(4)} & \colhead{(5)} & \colhead{(6)} & \colhead{(7)} & \colhead{(8)} & \colhead{(9)} & \colhead{(10)} & \colhead{(11)}}
\startdata
{\it AS 205}  & 0.029 & 0.9 & 46  & 19.6 & 0.11 & 0.14 & 25 & 165 & 2.1 & 3.7 \\
Elias 24      & 0.117 & 0.9 & 127 & 8.6  & 0.03 & 0.16 & 24 & 50  & 2.0 & 1.5 \\
{\it GSS 39}  & 0.143 & 0.7 & 198 & 7.3  & 0.08 & 0.07 & 60 & 110 & 1.9 & 32  \\
{\it AS 209}  & 0.028 & 0.4 & 126 & 13.3 & 0.10 & 0.09 & 38 & 86  & 1.7 & 2.4 \\
{\it DoAr 25} & 0.136 & 0.9 & 80  & 6.7  & 0.15 & 0.06 & 59 & 112 & 1.9 & 9.2 \\
{\it WaOph 6} & 0.077 & 1.0 & 153 & 4.4  & 0.06 & 0.12 & 39 & 171 & 1.8 & 1.8 \\
{\it VSSG 1}  & 0.029 & 0.8 & 33  & 9.7  & 0.08 & 0.10 & 53 & 165 & 1.8 & 12  \\
SR 4          & 0.004 & 0.8 & 20  & 19.9 & 0.23 & 0.07 & 50 & 39  & 1.9 & 8.8 \\
SR 13         & 0.012 & 1.0 & 26  & 12.6 & 0.07 & 0.04 & 32 & 42  & 2.0 & 1.9 \\
WSB 52        & 0.007 & 1.1 & 26  & 14.5 & 0.19 & 0.06 & 46 & 120 & 1.9 & 6.6 \\
DoAr 33       & 0.007 & 1.1 & 38  & 2.7  & 0.06 & 0.07 & 43 & 102 & 1.9 & 22  \\
WL 18         & 0.011 & 0.8 & 14  & 5.6  & 0.07 & 0.04 & 48 & 115 & 1.9 & 13  \\
\enddata
\tablecomments{Col.~(1): Disk name (those in italics were modeled in Paper I).  
Col.~(2): Disk mass assuming a 100:1 gas-to-dust mass ratio.  Col.~(3): Radial 
surface density gradient.  Col.~(4): Characteristic radius.  Col.~(5): Scale 
height at 100\,AU.  Col.~(6): Radial scale height gradient.  Col.~(7): Fixed 
inner radius.  Col.~(8): Fixed inclination.  Col.~(9): Fixed major axis 
position angle.  Col.~(10): Reduced $\chi^2$ statistic comparing the model fit 
with the continuum visibilities alone.  Col.~(11): Same as Col.~(10), but for 
the SED alone.}
\end{deluxetable}

\begin{deluxetable}{lccccccc|cc|cc}
\tablecolumns{12}
\tablewidth{0pt}
\tablecaption{Disk Structure Model Parameters: Central Cavity Cases\label{trans_table}}
\tablehead{
\colhead{Name} & \colhead{$M_d$} & \colhead{$\gamma$} & \colhead{$R_c$} & \colhead{$H_{100}$} & \colhead{$\psi$} & \colhead{$R_{\rm cav}$} & \colhead{$\delta_{\rm cav}$} & \colhead{$i$} & \colhead{PA} & \colhead{$\tilde{\chi}^2_{\rm vis}$} & \colhead{$\tilde{\chi}^2_{\rm sed}$} \\
\colhead{} & \colhead{[M$_{\odot}$]} & \colhead{} & \colhead{[AU]} & \colhead{[AU]} & \colhead{} & \colhead{[AU]} & \colhead{} & \colhead{[\degr]} & \colhead{[\degr]} & \colhead{} & \colhead{} \\
\colhead{(1)} & \colhead{(2)} & \colhead{(3)} & \colhead{(4)} & \colhead{(5)} & \colhead{(6)} & \colhead{(7)} & \colhead{(8)} & \colhead{(9)} & \colhead{(10)} & \colhead{(11)} & \colhead{(12)}}
\startdata
SR 24 S       & 0.042 & 0.8 & 40  & 4.0  & 0.02 & 32 & 0.0001 & 57 & 31  & 2.1 & 11  \\
{\it SR 21}   & 0.005 & 0.9 & 17  & 7.7  & 0.26 & 37 & 0.005  & 22 & 110 & 1.7 & 7.2 \\
{\it WSB 60}  & 0.021 & 0.8 & 31  & 11.0 & 0.13 & 20 & 0.01   & 25 & 117 & 1.8 & 3.0 \\
{\it DoAr 44} & 0.017 & 1.0 & 80  & 3.5  & 0.04 & 33 & 0.0001 & 45 & 75  & 1.8 & \nodata \\
\enddata
\tablecomments{Cols.~(1-6): Same as for Table \ref{structure_table}.  Col.~(7): 
The cavity radius, marking the outer edge of the diminished inner disk 
densities (see Paper I or the discussion on SR 24 S in \S 4.2).  Col.~(8): The 
density reduction scaling factor inside the radius $R_{\rm cav}$.  
Cols.~(9-12): Same as for Table \ref{structure_table} cols.~(8-11).}
\end{deluxetable}

\clearpage

\begin{deluxetable}{lcclc}
\tablecolumns{5}
\tablewidth{0pt}
\tablecaption{Viscous Disk Properties\label{viscous_table}}
\tablehead{
\colhead{Name} & \colhead{$\dot{M}_{\ast}$} & \colhead{$R_t$} & \colhead{$\alpha$} & \colhead{ref} \\ 
\colhead{} & \colhead{[M$_{\odot}$ yr$^{-1}$]} & \colhead{[AU]} & \colhead{} & \colhead{}          \\ 
\colhead{(1)} & \colhead{(2)} & \colhead{(3)} & \colhead{(4)} & \colhead{(5)}}
\startdata
{\it AS 205}  & $8\times10^{-8}$         & 23 & 0.005       & 1       \\
Elias 24      & $2\times10^{-7}$         & 62 & 0.03        & 2       \\
{\it GSS 39}  & $7\times10^{-8}$         & 95 & 0.03        & 2       \\
{\it AS 209}  & $9\times10^{-8}$         & 61 & 0.08        & 3       \\
{\it DoAr 25} & $3\times10^{-9}$         & 39 & 0.0005      & 4       \\
{\it WaOph 6} & $1\times10^{-7}$         & 78 & 0.05        & 5       \\
{\it VSSG 1}  & $1\times10^{-7}$         & 16 & 0.02        & 2       \\
SR 4          & $6\times10^{-8}$         & 10 & 0.02        & 2       \\
SR 13         & $3\times10^{-9}$         & 13 & 0.0005      & 2       \\
WSB 52        & $4\times10^{-9}$         & 13 & 0.001       & 2       \\
DoAr 33       & $3\times10^{-10}$        & 20 & 0.0006      & 6       \\
WL 18         & $9\times10^{-9}$         &  7 & 0.003       & 2       \\
\hline
SR 24 S       & $3\times10^{-8}$         & 19 & (0.006)     & 2       \\
{\it SR 21}   & $< 2\times10^{-9}$       &  9 & ($<$0.0009) & 2       \\
{\it WSB 60}  & $1\times10^{-9}$         & 15 & (0.0002)    & 2       \\
{\it DoAr 44} & $9\times10^{-9}$         & 40 & (0.01)      & 7       \\
\enddata
\tablecomments{Col.~(1): Disk name (those in italics were modeled in Paper I).  
Col.~(2): Accretion rate.  Col.~(3): Radius of mass flow reversal.  Col.~(4): 
Viscosity coefficient.  Col.~(5): References for $\dot{M}_{\ast}$: [1] - 
\citet{prato03}, [2] - \citet{natta06}, [3] - \citet{johnskrull00}, [4] - 
\citet{luhman99}, [5] - \citet{eisner05}, [6] - \citet{cieza10}, [7] - 
\citet{espaillat10}.}
\end{deluxetable}

\clearpage

\clearpage

\clearpage

\clearpage

\clearpage

\clearpage

\clearpage

\clearpage

\clearpage

\clearpage

\clearpage

\begin{thebibliography}{}
\bibitem[Alexander \& Armitage(2007)]{alexander07} Alexander, R. D., \& Armitage, P. J. 2007, \mnras, 375, 500
\bibitem[Alibert et al.(2005)]{alibert05} Alibert, Y., Mordasini, C., Benz, W., \& Winisdoerffer, C. 2005, \aap, 434, 343
\bibitem[Andr{\'{e}} \& Montmerle(1994)]{andre94} Andr{\'{e}}, P., \& Montmerle, T. 1994, \apj, 420, 837
\bibitem[Andrews \& Williams(2005a)]{aw05a} Andrews, S. M., \& Williams, J. P. 2005, \apj, 619, L175 (2005a)
\bibitem[Andrews \& Williams(2005b)]{aw05} --------- 2005, \apj, 635, 1134 (2005b)
\bibitem[Andrews \& Williams(2007a)]{aw07} --------- 2007, \apj, 659, 705 (2007a)
\bibitem[Andrews \& Williams(2007b)]{aw07b} --------- 2007, \apj, 671, 1800 (2007b)
\bibitem[Andrews et al.(2009)]{andrews09} Andrews, S. M., Wilner, D. J., Hughes, A. M., Qi, C., \& Dullemond, C. P. 2009, \apj, 700, 1502 (Paper I)
\bibitem[Barranco \& Goodman(1998)]{barranco98} Barranco, J. A., \& Goodman, A. A. 1998, \apj, 504, 207
\bibitem[Barsony et al.(2005)]{barsony05} Barsony, M., Ressler, M. E., \& Marsh, K. A. 2005, \apj, 630, 381
\bibitem[Beckwith et al.(1990)]{beckwith90} Beckwith, S. V. W., Sargent, A. I., Chini, R. S., \& G\"{u}sten, R. 1990, \aj, 99, 924
\bibitem[Boley et al.(2006)]{boley06} Boley, A. C., Mej{\'{\i}}a, A. C., Durisen, R. H., Cai, K., Pickett, M. K., \& D'Alessio, P. 2006, \apj, 651, 517
\bibitem[Boley(2009)]{boley09} Boley, A. C. 2009, \apj, 695, L53
\bibitem[Bontemps et al.(2001)]{bontemps01} Bontemps, S., et al. 2001, \aap, 372, 173
\bibitem[Boss(1997)]{boss97} Boss, A. P. 1997, Science, 276, 1836
\bibitem[Bouvier \& Appenzeller(1992)]{bouvier92} Bouvier, J., \& Appenzeller, I. 1992, \aaps, 92, 481
\bibitem[Brown et al.(2008)]{brown08} Brown, J. M., Blake, G. A., Qi, C., Dullemond, C. P., \& Wilner, D. J. 2008, \apj, 675, L109
\bibitem[Brown et al.(2009)]{brown09} Brown, J. M., Blake, G. A., Qi, C., Dullemond, C. P., Wilner, D. J., \& Williams, J. P. 2009, \apj, 704, 496
\bibitem[Caselli et al.(2002)]{caselli02} Caselli, P., Benson, P. J., Myers, P. C., \& Tafalla, M. 2002, \apj, 572, 238
\bibitem[Chambers \& Cassen(2002)]{chambers02} Chambers, J. E., \& Cassen, P. 2002, Meteoritics \& Planetary Science, 37, 1523
\bibitem[Chiang \& Goldreich(1997)]{chiang97} Chiang, E. I., \& Goldreich, P. 1997, \apj, 490, 368
\bibitem[Ciesla(2007)]{ciesla07} Ciesla, F. J. 2007, \apj, 654, L159
\bibitem[Cieza et al.(2010)]{cieza10} Cieza, L. A., et al. 2010, \apj, 712, 925
\bibitem[Crida(2009)]{crida09} Crida, A. 2009, \apj, 698, 606
\bibitem[Cutri et al.(2003)]{cutri03} Cutri, R. M., et al. 2003, 2MASS All-Sky Point Source Catalog (Pasadena: IPAC)
\bibitem[D'Alessio et al.(2005)]{dalessio05} D'Alessio, P. et al. 2005, \apj, 621, 461
\bibitem[Dartois et al.(2003)]{dartois03} Dartois, E., Dutrey, A., \& Guilloteau, S. 2003, \aap, 399, 773
\bibitem[de Geus et al.(1989)]{degeus89} de Geus, E. J., de Zeeuw, P. T., \& Lub, J. 1989, \aap, 216, 44
\bibitem[Dib et al.(2010)]{dib10} Dib, S., Hennebelle, P., Pineda, J. E., Csengeri, T., Bontemps, S., Audit, E., \& Goodman, A. A. 2010, \apj, in press (arXiv:1003.5118)
\bibitem[Draine \& Lee(1984)]{draine84} Draine, B. T., \& Lee, H. M. 1984, \apj, 285, 89
\bibitem[Dullemond \& Dominik(2004)]{dullemond04a} Dullemond, C. P., \& Dominik, C. 2004, \aap, 417, 159 
\bibitem[Dullemond \& Dominik(2005)]{dullemond05} --------- 2005, \aap, 434, 971
\bibitem[Durisen et al.(2007)]{durisen07} Durisen, R. H., Boss, A. P., Mayer, R. L., Nelson, A. F., Quinn, T., \& Rice, W. K. M. 2007, in Protostars \& Planets V, eds. B.~Reipurth, D.~Jewitt, \& K.~Keil (Tucson: Univ.~Arizona Press), 607
\bibitem[Eisner et al.(2005)]{eisner05} Eisner, J. A., Hillenbrand, L. A., White, R. J., Akeson, R. L., \& Sargent, A. I. 2005, \apj, 623, 952
\bibitem[Espaillat et al.(2007)]{espaillat07} Espaillat, C., et al. 2007, \apj, 670, L135
\bibitem[Espaillat et al.(2010)]{espaillat10} --------- 2010, \apj, in press (arXiv:1005.2365)
\bibitem[Evans et al.(2003)]{evans03} Evans, N. J., et al. 2003, \pasp, 115, 965
\bibitem[Fleming \& Stone(2003)]{fleming03} Fleming, T., \& Stone, J. M. 2003, \apj, 585, 908
\bibitem[Fromang et al.(2007)]{fromang07} Fromang, S., Papaloizou, J., Lesur, G., \& Heinemann, T. 2007, \aap, 476, 1123
\bibitem[Furlan et al.(2009)]{furlan09} Furlan, E., et al. 2009, \apj, 703, 1964
\bibitem[Garaud(2007)]{garaud07} Garaud, P. 2007, \apj, 671, 2091
\bibitem[Ghez et al.(1993)]{ghez93} Ghez, A. M., Neugebauer, G., \& Matthews, K. 1993, \aj, 106, 2005
\bibitem[Goodman et al.(1993)]{goodman93} Goodman, A. A., Benson, P., Fuller, G., \& Myers, P. 1993, \apj, 406, 528
\bibitem[Gorti \& Hollenbach(2009)]{gorti09a} Gorti, U., \& Hollenbach, D. 2009, \apj, 690, 1539
\bibitem[Gorti et al.(2009)]{gorti09b} Gorti, U., Dullemond, C. P., \& Hollenbach, D. 2009, \apj, 705, 1237
\bibitem[Gullbring \& Calvet(1998)]{gullbring98} Gullbring, E., \& Calvet, N. 1998, \apj, 509, 802
\bibitem[Gullbring et al.(2000)]{gullbring00} Gullbring, E., Calvet, N., Muzerolle, J., \& Hartmann, L. 2000, \apj, 544, 927
\bibitem[Hartmann et al.(1998)]{hartmann98} Hartmann, L., Calvet, N., Gullbring, E., \& D'Alessio, P. 1998, \apj, 495, 385
\bibitem[Hawley et al.(1995)]{hawley95} Hawley, J. F., Gammie, C. F., \& Balbus, S. A. 1995, \apj, 440, 742
\bibitem[Herbig \& Bell(1988)]{herbig88} Herbig, G. H., \& Bell, K. R. 1988, in Third Catalog of Emission Line Stars of the Orion Population (Santa Cruz: Lick Obs.)
\bibitem[Herbst et al.(1994)]{herbst94} Herbst, W., Herbst, D. K., \& Grossman, E. J. 1994, \aj, 108, 1906
\bibitem[Hillenbrand(2009)]{hillenbrand09} Hillenbrand, L. A. 2009, in IAU Symp.~258, The Age of Stars, ed. E. Mamajek, et al. (Cambridge: Cambridge University Press), 81
\bibitem[Ho et al.(2004)]{ho04} Ho, P. T. P., Moran, J. M., \& Lo, K. Y. 2004, \apj, 616, L1
\bibitem[Hubickyj et al.(2005)]{hubickyj05} Hubickyj, O., Bodenheimer, P., \& Lissauer, J. J. 2005, Icarus, 179, 415
\bibitem[Hueso \& Guillot(2005)]{hueso05} Hueso, R., \& Guillot, T. 2005, \aap, 442, 703
\bibitem[Hughes et al.(2007)]{hughes07} Hughes, A. M., Wilner, D. J., Calvet, N., D'Alessio, P., Claussen, M. J., \& Hogerheijde, M. R. 2007, \apj, 664, 536
\bibitem[Hughes et al.(2008)]{hughes08} Hughes, A. M., Wilner, D. J., Qi, C., \& Hogerheijde, M. R. 2008, \apj, 678, 1119
\bibitem[Hughes et al.(2009)]{hughes09} Hughes, A. M., et al. 2009, \apj, 698, 131
\bibitem[Ida \& Lin(2004)]{ida04} Ida, S., \& Lin, D. N. C. 2004, \apj, 604, 388
\bibitem[Isella et al.(2009)]{isella09} Isella, A., Carpenter, J. M., \& Sargent, A. I. 2009, \apj, 701, 260
\bibitem[Isella et al.(2010)]{isella10} --------- 2010, \apj, 714, 1746
\bibitem[Jensen \& Mathieu(1997)]{jensen97} Jensen, E. L. N., \& Mathieu, R. D. 1997, \aj, 114, 301
\bibitem[Johns-Krull et al.(2000)]{johnskrull00} Johns-Krull, C. M., Valenti, J. A., \& Linsky, J. L. 2000, \apj, 539, 815
\bibitem[Kitamura et al.(2002)]{kitamura02} Kitamura, Y., Momose, M., Yokogawa, S., Kawabe, R., Tamura, M., \& Ida, S. 2002, \apj, 581, 357
\bibitem[Kokubo \& Ida(2002)]{kokubo02} Kokubo, E., \& Ida, S. 2002, \apj, 581, 666
\bibitem[Knude \& H{\o}g(1998)]{knude98} Knude, J., \& H{\o}g, E. 1998, \aap, 338, 897
\bibitem[Loinard et al.(2008)]{loinard08} Loinard, L., Torres, R. M., Mioduszewski, A. J., \& Rodr{\'{\i}}guez, L. F. 2008, \apj, 675, L29
\bibitem[Lombardi et al.(2008)]{lombardi08} Lombardi, M., Lada, C. J., \& Alves, J. 2008, \aap, 480, L785
\bibitem[Lubow \& D'Angelo(2006)]{lubow06} Lubow, S. H., \& D'Angelo, G. 2006, \apj, 641, 526
\bibitem[Luhman \& Rieke(1999)]{luhman99} Luhman, K. L., \& Rieke, G. H. 1999, \apj, 525, 440
\bibitem[Lynden-Bell \& Pringle(1974)]{lyndenbell74} Lynden-Bell, D., \& Pringle, J. E. 1974, \mnras, 168, 603
\bibitem[Mayama et al.(2010)]{mayama10} Mayama, S., et al. 2010, Science, 327, 306
\bibitem[McCabe et al.(2006)]{mccabe06} McCabe, C., Ghez, A. M., Prato, L., Duchene, G., Fischer, R. S., \& Telesco, C. 2006, \apj, 636, 932
\bibitem[McCaughrean \& O'Dell(1996)]{mccaughrean96} McCaughrean, M. J., \& O'Dell, C. R. 1996, \aj, 111, 1977
\bibitem[Mordasini et al.(2009)]{mordasini09} Mordasini, C., Alibert, Y., \& Benz, W. 2009, \aap, 501, 1139
\bibitem[Motte et al.(1998)(1998)]{motte98} Motte, F., Andr{\'{e}}, P., \& Neri, R. 1998, \aap, 336, 150
\bibitem[Muzerolle et al.(1998a)]{muzerolle98a} Muzerolle, J., Calvet, N., \& Hartmann, L. 1998, \apj, 492, 743
\bibitem[Muzerolle et al.(1998b)]{muzerolle98b} Muzerolle, J., Hartmann, L., \& Calvet, N. 1998, \aj, 116, 2965 (1998b)
\bibitem[Natta et al.(2006)]{natta06} Natta, A., Testi, L., \& Randich, S. 2006, \aap, 452, 245
\bibitem[Najita et al.(2007)]{najita07} Najita, J. R., Strom, S. E., \& Muzerolle, J. 2007, \mnras, 378, 369 
\bibitem[N\"{u}rnberger et al.(1998)]{nurnberger98} N\"{u}rnberger, D., Brandner, W., Yorke, H. W., \& Zinnecker, H. 1998, \aap, 330, 549
\bibitem[Padgett et al.(2008)]{padgett08} Padgett, D. C., et al. 2008, \apj, 672, 1013
\bibitem[Patience et al.(2008)]{patience08} Patience, J., Akeson, R. L., \& Jensen, E. L. N. 2008, \apj, 677, 616
\bibitem[Pi{\'{e}}tu et al.(2006)]{pietu06} Pi{\'{e}}tu, V., Dutrey, A., Guilloteau, S., Chapillon, E., \& Pety, J. 2006, \aap, 460, L43
\bibitem[Pi{\'{e}}tu et al.(2007)]{pietu07} Pi{\'{e}}tu, V., Dutrey, A., \& Guilloteau, S. 2007, \aap, 467, 163
\bibitem[Pollack et al.(1996)]{pollack96} Pollack, J. B., Hubickyj, O., Bodenheimer, P., Lissauer, J. J., Podolak, M., \& Greenzweig, Y. 1996, Icarus, 124, 62
\bibitem[Prato et al.(2003)]{prato03} Prato, L., Greene, T. P., \& Simon, M. 2003, \apj, 584, 853
\bibitem[Raymond et al.(2005)]{raymond05} Raymond, S. N., Quinn, T., \& Lunine, J. I. 2005, \apj, 632, 670
\bibitem[Reipurth \& Zinnecker(1993)]{reipurth93} Reipurth, B., \& Zinnecker, H. 1993, \aap, 278, 81
\bibitem[Schaefer et al.(2006)]{schaefer06} Schaefer, G. H., Simon, M., Beck, T. L., Nelan, E., \& Prato, L. 2006, \aj, 132, 2618
\bibitem[Shakura \& Sunyaev(1973)]{shakura73} Shakura, N. I., \& Sunyaev, R. A. 1973, \aap, 24, 337
\bibitem[Siess et al.(2000)]{siess00} Siess, L., Dufour, E., \& Forestini, M. 2000, \aap, 358, 593
\bibitem[Simon et al.(1995)]{simon95} Simon, M., et al. 1995, \apj, 443, 625
\bibitem[Stanke et al.(2006)]{stanke06} Stanke, T., Smith, M. D., Gredel, R., \& Kanzadyan, T. 2006, \aap, 447, 609
\bibitem[Stone et al.(1996)]{stone96} Stone, J. M., Hawley, J. F., Gammie, C. F., \& Balbus, S. A. 1996, \apj, 463, 656
\bibitem[Toomre(1964)]{toomre64} Toomre, A. 1964, \apj, 139, 1217
\bibitem[Weidenschilling(1977)]{weidenschilling77} Weidenschilling, S. J. 1977, \apss, 51, 153
\bibitem[Weingartner \& Draine(2001)]{weingartner01} Weingartner, J. C., \& Draine, B. T. 2001, \apj, 548, 296
\bibitem[Wilking et al.(2005)]{wilking05} Wilking, B. A., Meyer, M. R., Robinson, J. G., \& Greene, T. P. 2005, \aj, 130, 1733
\bibitem[Zhu et al.(2010)]{zhu10} Zhu, Z., Hartmann, L., \& Gammie, C. 2010, \apj, 713, 1143
\end{thebibliography}
\end{document}